\documentclass[reprint,prfluids,superscriptaddress,notitlepage,amsmath,amssymb,aps,onecolumn,floatfix,tightenlines,longbibliography,11pt]{revtex4-2}
\usepackage{graphicx}
\usepackage{color}
\usepackage{hyperref}
\usepackage{cleveref}
\usepackage{physics}

\renewcommand{\vec}[1]{\mathbf{#1}}
\newcommand{\xt}{(\vec{x},t)}
\newcommand{\uu}{\vec{u}}

\newcommand{\xone}{\vec{x}^{(1)}}
\newcommand{\xtwo}{\vec{x}^{(2)}}
\newcommand{\uone}{\vec{u}^{(1)}}
\newcommand{\utwo}{\vec{u}^{(2)}}
\newcommand{\aone}{\vec{a}^{(1)}}
\newcommand{\atwo}{\vec{a}^{(2)}}
\newcommand{\average}[1]{\left\langle #1 \right\rangle}

\newcommand{\Rey}{\mathrm{Re}}
\newcommand{\Fr}{\mathrm{Fr}}
\newcommand{\Ro}{\mathrm{Ro}}
\newcommand{\Pran}{\mathrm{Pr}}
\newcommand{\RIB}{R_{\mathrm{IB}}}

\begin{document}

\title{Multi-Particle Dispersion in Rotating-Stratified Turbulent Flows}
\date{\today}

\author{Sebastian Gallon}
\affiliation{Univ Lyon, ENS de Lyon, CNRS, Laboratoire de Physique, F-69342 Lyon, France}
\author{Fabio Feraco}
\affiliation{CNRS, École Centrale de Lyon, INSA de Lyon, Univ Claude Bernard Lyon 1, Laboratoire de Mécanique des Fluides et d’Acoustique - UMR 5509, F-69134 Écully, France}
\affiliation{Dipartimento di Fisica, Universit\`a della Calabria, Rende I-87036, Italy}
\author{Raffaele Marino}
\affiliation{CNRS, École Centrale de Lyon, INSA de Lyon, Univ Claude Bernard Lyon 1, Laboratoire de Mécanique des Fluides et d’Acoustique - UMR 5509, F-69134 Écully, France}
\author{Alain Pumir}
\affiliation{Univ Lyon, ENS de Lyon, CNRS, Laboratoire de Physique, F-69342 Lyon, France}
\affiliation{Max Planck Institute for Dynamics and Self-Organization, 37077 Göttingen, Germany}
\date{\today}

\begin{abstract}
The transport of matter by turbulent flows plays an important role, in particular in a geophysical context. Here, we study the relative movement of groups of two (pairs) and four (tetrahedra) Lagrangian particles using direct numerical simulations of the stably-stratified Boussinsesq equations, with Brunt-V\"ais\"al\"a frequency $N$ and Coriolis parameter $f$. We cover regimes close to homogeneous isotropic turbulence, to flows dominated by stratification and rotation, keeping fixed the ratio $N/f = 5$. The flows studied are anisotropic, so the relative motion between two particles depends not only on the initial separation between the particles, but also on their orientation with respect to the vertical axis. In all cases considered, we demonstrate that the relative particle motion differs depending on whether dispersion is considered forward or backwards in time, although the asymmetry becomes less pronounced when stratification and rotation increase. On the other hand, the strong fluctuations in the dispersion between two particles become more extreme when $N$ and $f$ increase. We also find evidence for the formation of shear layers, which become more pronounced as $N$ and $f$ become larger. Finally, we show that the irreversibility on the dispersion of a set of particles forming initially a regular tetrahedron becomes weaker when the influence of stratification and rotation increase, a property that we relate to that of the perceived rate-of-strain tensor.
\end{abstract}
\maketitle

\section{Introduction}
The dispersion of material substances in turbulent fluid flows notoriously plays an important role in many contexts, in particular in geophysics~\cite{pedloskyGeophysical1979,weissTransport2008}. One example of particular interest is the transport of biomatter in the oceans~\cite{guastoFluid2012}. Phytoplankton blooms, which play both a crucial role in the maritime eco-system~\cite{diazSeasonal2021,salleeSummertime2021} and for the exchange of greenhouse gasses between the atmosphere and the oceans~\cite{wihsgottObservations2019}, are deeply affected by mixing in the upper and middle ocean. A detailed understanding of the mixing in the oceans, however, is hindered by the complexity of the underlying turbulent flows, with a wide range of scales, from the largest turbulent structures, of the order of tens of kilometers, down to the dissipative scales, of the order of a few the millimeters~\cite{thorpeIntroduction2007}. In addition, these flows are stratified, and subject to the rotation of the Earth. These features renders the modeling of the mixing process very challenging. In particular, direct numerical simulations (DNS) of such flows can hardly reproduce full oceanic dynamics, though realistic regimes in terms of at least some of the governing parameters have been attained using very large grids \cite{marino_15,rosenberg_15}.\par
A natural framework to study dispersion properties is to investigate the motion of neutrally buoyant tracer particles~\cite{moninStatistical1975}. This approach has been used to understand the fundamental properties of turbulent flows, particularly in the case of homogeneous and isotropic turbulence (HIT)~\cite{sawfordTurbulent2001,yeungLagrangian2002,toschiLagrangian2009}. Although one of the central tenets of turbulence theory is that the framework of HIT describes the small scale motion of turbulent flows in the limit of very large Reynolds numbers~\cite{moninStatistical1975,frischTurbulence1995}, it remains of great fundamental and practical interest to understand the influence of stratification and rotation on transport properties of the flow. In fact, in the oceans, the influence of stratification remains significant, even in the most turbulent regions. A particular consequence of stratification is the inhibition of vertical velocity fluctuations, and therefore the exchange of heat and material substances between the upper and lower ocean~\cite{pedloskyGeophysical1979,thorpeIntroduction2007}.
Furthermore, rotation also plays a role in favoring horizontal over vertical motions by promoting bi-dimensional modes of the flow~\citep{Marino_2013}, although its influence is  generally weaker in the setup considered here where $N/f = 5$. \par
In the present work, we study numerically dispersion in an idealized model of geophysical turbulence, namely in the Boussinesq equations with a linear, stable density gradient and solid body rotation. The relative strength of rotation and stratification is chosen specifically for its oceanographic relevance~\cite{garabatoWidespread2004,nikurashinRoutes2013}. In particular, we extend the recent work of~\cite{buariaSingleparticle2020}, which was focusing on the motion of single particles, to the relative dispersion of pairs and groups of four particles. By including the role of both stratification and rotation simultaneously, our work also extends earlier studies for flows with pure rotation~\cite{nasoMultiscale2019,polancoMultiparticle2023} or with pure stratification~\cite{vanaartrijkSingleparticle2008}. \par
Groups of particles in a complex flow are not only separating, but due to both large scale structures and small scale fluctuations, distant particles can also come close together. This questions is of relevance for mixing related phenomena~\cite{sawfordTurbulent2001}. A natural framework to understand this motion is to study backwards dispersion. To this end, we consider pairs of particles that are separated by a small distance at time $t=0$ and follow their trajectories backwards in time. Moreover, the study of differences between forwards and backwards dispersion for both single particles and for groups of particles has been shown useful to highlight the irreversible nature of turbulent flows~\cite{xuFlight2014,juchaTimereversalsymmetry2014,braggIrreversibility2018,cheminetEulerian2022,gallonLagrangian2024a}. In our simplified problem, we show that the irreversibility becomes weaker when stratification and rotation are increased. \par
Whilst the transport of one particle is governed directly by the velocity field at its position, the relative dispersion of groups of particles is governed by velocity differences and consequently probe the intermittent nature of the flows~\cite{scatamacchiaExtreme2012,bitaneGeometry2013,buariaCharacteristics2015}. In particular, the large fluctuations in the velocity gradient~\cite{buariaExtreme2019,buariaVorticityStrain2022} may induce strong relative velocities of particles very close to each other. This provides a natural explanation for the extreme fluctuations reported in numerical studies of the separation of two nearby particles in HIT~\cite{scatamacchiaExtreme2012}. One of the unexpected findings of the present study is that the strong fluctuations in particle dispersions are enhanced by the presence of rotation and stratification, i.e., pair dispersion appears to be more intermittent when increasing the strength of rotation and stratification than a comparable HIT flow. \par
Moreover, we notice that in an anisotropic flow, the separation of two particles depends not only on the initial distance between the particles, but also on their alignment with respect to the preferred direction of the flow i.e. the vertical in our problem. By documenting the dependence of the dispersion on the orientation, we demonstrate that the flow develops horizontal shear layers, the more so as stratification and rotation get stronger. In this sense, our results shed new light on the structure of these flows.\par 
Finally, we study the effect of rotation and stratification on the deformation of fluid volumes by considering how groups of four originally equidistant particles separate. Whereas the dispersion of pairs of particles is completely governed by velocity and buoyancy differences, the dynamics of larger groups of particles, is also sensitive to the complex three-dimensional structure of the flow. Specifically, we focus on differences of the deformation forwards and backwards in time, highlighting the irreversibility of the flow, similarly as the previously studied case of HIT~\cite{juchaTimereversalsymmetry2014}. Moreover, we find that at intermediate and larger times, such tetrahedra tend to dealign with the vertical axes. \par
Our work is organized as follows. In \cref{sec:background}, we describe our DNS, and introduce the key concepts used to analyze our simulations. The results are presented in \cref{sec:Results}, which focuses first on pair dispersion (\cref{subsec:pair_dispersion}), and then on dispersion of more complex objects, namely tetrads (\cref{subsubsec:volume_def}). Our conclusions are presented in \cref{sec:conclusion}.

\section{Background and Methodology}
\label{sec:background}
\subsection{Direct numerical simulations and physical background}\label{ssc:dns}
We perform direct numerical simulations of the Boussinesq equations in a triply periodic domain in a rotating frame with solid body rotation rate $\Omega$ (the Coriolis frequency is defined by $f = 2\Omega$). The fluid is stably stratified, with a linear mean density profile $\rho_0(z) = \bar{\rho} - \gamma z$ with constants $\rho$ and $\gamma > 0$ and Brunt-V\"ais\"al\"a frequency $N = \sqrt{\gamma g/\rho_0}$. The equations of motion for the velocity field $\vec{u}$ and the density fluctuation field in the dimension of a velocity $\theta\xt =N[\rho\xt - \rho_0\xt]/\gamma \qc$ are given by 
\begin{align}
    \label{eq:Bous_u}
    \partial_t \vec{u}  + \vec{u} \cdot \boldsymbol{\nabla} \vec{u} &= 
    -\grad p - f \vec{e}_z  \cross \vec{u}  - N \theta \vec{e}_z 
    + \nu \laplacian \vec{u} + \vec{\Phi} \qq*{,}  \\
    \label{eq:Bous_T}
    \partial_t \theta + \vec{u} \cdot \grad \theta &= 
    N w + \kappa \laplacian \theta \qq*{,} \\
    \div \vec{u} &= 0 \qq*{,} \label{eq:Bous_div}
    \end{align}
where $p\xt$ denotes the reduced pressure, $\nu$ the kinematic viscosity and $\kappa$ the thermal diffusivity. Here, we choose $\Pran = \nu/\kappa = 1$. \Cref{eq:Bous_u,eq:Bous_T,eq:Bous_div} are integrated in a triply periodic domain with $M^3$ Fourier modes, where $M=512$ (runs A0 - A4) and $M=1024$ (runs B0 - B4). We use the pseudo-spectral solver GHOST (Geophysical High-Order Suite for Turbulence), with a second-order explicit Runge-Kutta time-stepping scheme~\cite{mininniHybrid2011,Rosenberg_2020}. The time-step $\mathrm{d}t$ was chosen to be sufficiently small to resolve the fastest propagating waves. To maintain a stationary flow, we apply a stochastic large-scale force $\vec{\Phi}\xt$. The forcing is isotropic and applied in a wave number band $2 \leq \abs{\vec{k}} \leq 3$ to the velocity field. The resulting characteristic length scale of the forcing reads $L_\vec{\Phi} = 2 \pi / 2.5$, here considered as the integral scale of the system.

To quantify the relative strengths of turbulence, stratification and rotation, we consider the dimensionless Reynolds ($\Rey$), Froude ($\Fr$) and Rossby ($\Ro$) numbers, defined as
\begin{equation}
    \Rey = \frac{u' L_\Phi}{\nu} \qq{,} \Fr = \frac{u'}{N L_\Phi} \qq{,} \Ro = \frac{u'}{f L_\Phi} \qc
\end{equation}
the characteristic large-scale velocity being defined as $u' = \langle \abs{\vec{u}}^2 \rangle^{1/2}$. In this work, we consider a series of runs with varying $\Rey$, $\Fr$ and $\Ro$, where we fix the relative strength of the rotation and stratification by setting the ratio $N/f = \Ro/\Fr = 5$, representative of the Southern Ocean~\cite{garabatoWidespread2004,nikurashinRoutes2013}.

In the framework considered, we observe a forward energy cascade as a result of a large-scale energy injection, at rate $\varepsilon_\vec{\Phi} = \langle \uu \cdot \vec{\Phi} \rangle$. At intermediate scales, part of the kinetic energy is transformed
into potential energy at the rate $\varepsilon_\chi = N \langle \theta w \rangle$. The remaining part of the kinetic energy is transferred to smaller scale, until it gets dissipated at rate $\varepsilon_\nu = \nu \langle\abs{\grad \vec{u}}^2 \rangle$. The characteristic scale for energy dissipation is the Kolmogorov scale $\eta = (\nu^3/\varepsilon_\nu)^{1/4}$ \cite{davidsonTurbulence2013}. The corresponding typical time scale for dissipation is the Kolmogorov time scale $\tau_\eta = (\nu/\varepsilon_\nu)^{1/2}$. To ensure that the flow is well resolved, we kept $k_\mathrm{max}\eta \gtrsim 1.4$, where $k_\mathrm{max} = M/3$ is the maximal wave number resolved in the simulation.  \par
Stratified fluid turbulence is characterized by the formation of horizontal layers with a typical vertical size $L_B = u^\prime / N$, making the flow anisotropic \cite{Marino_2014}. The length scale below which the flow recovers isotropy is called the Ozmidov scale $\ell_O = (\varepsilon_\nu/N^3)^{1/2}$~\cite{rileyStratified2008}. The equivalent rotational length scale is the Zeman scale $\ell_Z = (\varepsilon_\nu/\Omega^3)^{1/2}$~\cite{zemanNote1994,Pouquet_2019}. In the configuration considered here, we have $\ell_Z \approx 3.95 \, \ell_O$. 

The previously considered dimensionless number are large-scale quantities. To compare the relative strength of stratification and dissipation, we consider the buoyancy Reynolds number $\RIB = \varepsilon_\nu/(\nu N^2) $~\cite{davidsonTurbulence2013} (sometimes referred to as the buoyancy interaction parameter \cite{Pouquet_2019}). It can be seen as both a ratio of length scales and a ratio of time scales, as $\RIB = (\ell_O/\eta)^{4/3} = (\tau_\eta N)^{-2}$, with the typical time scale for buoyancy oscillations $\sim 1/N$. For a review and a comparison to the related $R_\mathcal{B}= \Re \Fr^2$ see \cite{iveyDensity2008}. A transition from eddy-dominated to wave-dominated flows has been recently observed at $\RIB \approx 1$ for both the motion of single particles~\cite{buariaSingleparticle2020} and pairs of particles~\cite{gallonLagrangian2024a}. 

During the simulations, we integrated the trajectories of $\sim 10^6$ naturally buoyant tracer particles. 
The parameters of the runs are indicated in \cref{tab:param}.

\begin{table}[tb] 
\centering
\caption{Parameters of the runs in dimensionless units.
    $N$ Brunt-V\"ais\"ail\"a frequency,
    $\varepsilon_\nu$ kinetic energy dissipation rate (averaged over particle integration time), 
    $\mathrm{Fr}$ Froude number, 
    $\mathrm{Re}$ Reynolds number, 
    $R_{\mathrm{IB}}$ buoyancy Reynolds number,
    $\eta$ Kolmogorov scale, 
    $\ell_O$ Ozmidov scale, 
    $L_B$ buoyancy scale. All quantities are rounded to three significant digits. We chose ran runs A0 - A4 a grid resolution of $M=512$ and runs B0 - B4 at $M = 1024$. The Reynolds number based on the Taylor scale for the HIT runs read $\mathrm{Re}_\lambda = 110$ (run A0) and $\mathrm{Re}_\lambda = 160$ (run B0). The kinematic viscosity was set to $\nu = 1.5 \cdot 10^{-3}$ for run A0, $\nu = 1 \cdot 10^{-3}$ for runs A1-A4 and $\nu = 2.1 \cdot 10^{-4}$ for runs B0-B4.} \label{tab:param}
\begin{ruledtabular}
\begin{tabular}{lllllllllll}
Run & A0 & A1 & A2 & A3 & A4 & B0 & B1 & B2 & B3 & B4 \\
$N$ & 0 & 2.95 & 4.92 & 7.37 & 14.7 & 0 & 1.18 & 2.95 & 7.37 & 14.7 \\
$\varepsilon_\nu$ & 0.375 & 0.206 & 0.156 & 0.123 & 0.0237 & 0.00784 & 0.00932 & 0.00715 & 0.0136 & 0.0344 \\
$\mathrm{Fr}$ & $\infty$ & 0.169 & 0.105 & 0.0777 & 0.0410 & $\infty$ & 0.155 & 0.0823 & 0.0599 & 0.0356 \\
$\mathrm{Re}$ & 2380 & 3140 & 3270 & 3620 & 3820 & 4790 & 5510 & 7300 & 13300 & 15800 \\
$R_{\mathrm{IB}}$ & $\infty$ & 23.8 & 6.47 & 2.26 & 0.109 & $\infty$ & 31.9 & 3.92 & 1.19 & 0.755 \\
$\eta$ & 0.00974 & 0.00834 & 0.00894 & 0.00950 & 0.0143 & 0.00586 & 0.00561 & 0.00600 & 0.00511 & 0.00405 \\
$\ell_O / \eta$ & $\infty$ & 10.8 & 4.06 & 1.84 & 0.190 & $\infty$ & 13.4 & 2.78 & 1.14 & 0.810 \\
$L_B / \eta$ & $\infty$ & 50.8 & 29.6 & 20.6 & 7.20 & $\infty$ & 69.5 & 34.5 & 29.5 & 22.1 \\
\end{tabular}
\end{ruledtabular}
\end{table}
\subsection{Definitions for multi-particle Lagrangian statistics}
\subsubsection{Short time pair dispersion}
 Let us consider two particles with trajectories $\xone(t)$ and $\xtwo(t)$, initially separated by $\vec{r}(0) = \xtwo(0)-\xone(0) = \vec{r}(0)$. For very small times $t$, the motion of the particles are solely governed by the initial values of the relative velocity $\Delta \vec{u}(0) = \utwo(0) - \uone(0)$ and, in one order higher, by the initial relative acceleration $\Delta \vec{a}(0) = \atwo(0) - \aone(0)$. Then, it is customary to consider the squared change of separation 
 \begin{equation}
 \delta_{\vec{r}_0}r^2(t) = (\vec{r}(t) - \vec{r}(0))^2 \qq{.}
 \end{equation}
  A straightforward Taylor expansion and subsequently averaging over many pairs of particles with the same initial separation yields
\begin{equation}
    \langle \delta_{\vec{r}_0}r(t)^2 \rangle = \langle (\Delta_{\vec{r}_0}\vec{u})^2 \rangle t^2 + \langle (\Delta_{\vec{r}_0}\vec{u} \cdot \Delta_{\vec{r}_0}\vec{a}) \rangle t^3 + \mathrm{h.o.t.}\ \qq*{.} \label{eq:TaylorPair} 
\end{equation}
The time over which this expansion remains valid is estimated the time scale, where the quadratic term and the cubic term are equal~\cite{batchelorApplication1950,bitaneTime2012}
\begin{equation}\label{eq:tBatchelor}
t_B(r_0) := \abs{\frac{\langle{\qty(\Delta_{r_0} \vec{u})}^2 \rangle}{\langle \Delta_{r_0}\vec{u} \cdot  \Delta_{r_0} \vec{a} \rangle}} \qc
\end{equation}
which is also referred to as the Batchelor time. For homogenous isotropic turbulence, $t_B(r_0)$ can be approximated assuming $\langle\qty(\Delta_{r_0} \vec{u})^2 \rangle \propto \varepsilon_\nu^{2/3} r_0^{2/3}$ and $\langle {\Delta_{r_0}\vec{u}} \cdot \Delta_{r_0} \vec{a} \rangle = - 2 \varepsilon_\nu$ to $t_B(r_0) \propto \varepsilon_\nu^{-1/3}r_0^{2/3} = \tau_{r_0}$ that is the Kolmogorov time associated to a vortex of size $r_0$~\cite{juchaTimereversalsymmetry2014}. This approximation is not necessarily valid in the presence of rotation and stratification~\cite{gallonLagrangian2024a}, we therefore determine $t_B(r_0)$ by measuring both the numerator and denominator of \cref{eq:tBatchelor}. \par
At longer times $t \gg t_B(r_0)$, the Taylor expansion~\cref{eq:TaylorPair} fails to describe the pair dynamics. The scaling law is governed by the complex correlation structure of the flow and depends even for HIT sensitively on the Reynolds number and the initial value of the separation $r_0$~\cite{buariaCharacteristics2015}.
\subsubsection{Short time fluid volume deformation}
\label{subsubsec:volume_def}

We now study how larger groups of particles separate. Specifically, we are interested how initially regular shaped fluid volumes are deformed. To this end, we consider the smallest set of particles defining a volume, which is a set of four non-coplanar particles with trajectories $\vec{x}^{(1)}(t),\vec{x}^{(2)}(t),\vec{x}^{(3)}(t)$ and $\vec{x}^{(4)}(t)$. Such a set can be geometrically interpreted as a tetrahdron whose vertices are the particles. Specifically we consider groups of initally pairwise equidistant particles, or equivalently, particles that are forming initially regular tetrahedra. To study the deformation of such tetrahedra when its vertices are advected by the flow, we define the moment-of-inertia-like shape tensor~\cite{pumirTetrahedron2013}:
\begin{equation}
    G_{ij}(t):= \sum_{a=1}^4 \qty(x^{(a)}_{i}(t) - x^c_{i}(t))
    \qty(x^{(a)}_{j}(t) - x^c_{j}(t)) \qc
\end{equation}
where $x^{(a)}_{i}$ denotes the $i$-th component of $\vec{x}^{(a)}$ and   $\vec{x}^c$ is the center of mass $\vec{x}^c = \sum_{a=1}^4\vec{x}^{(a)}/4$. For four non-coplanar particles $\vec{G}$ is symmetric and positive definite and can thus be diagonalized. Its eigenvalues $g_1 \geq g_2 \geq g_3$ and corresponding eigenvectors $\vec{g}_1$, $\vec{g}_2$, $\vec{g}_3$ provide a qualitative description of the tetrahedron's geometry and orientation; given that $\sqrt{g_i}$ is the tetrahedron's extend in the $i$-th eigendirection $\vec{g}_i$~\cite{pumirTetrahedron2013}.
 For a regular tetrahedron, whose edges have the same length $R_0$, the eigenvalues read $g_1=g_2=g_3= R_0^2/2$. For the cases $g_1 \gg g_2 \approx g_3$ the tetrahedra has a needle-like shape, while for $g_1 \approx g_2 \gg g_3$ its shape is pancake-like. We note that these eigenvalues contain information on both the size and the shape of the tetrahedron. \par
 The linear size of such a tetrahedron can be estimated as the squared gyration radius 
\begin{equation}
    R_G^2(t) := \sum_{a=1}^4 \abs{\vec{x}^{(a)}-\vec{x}^c}^2 = \frac{1}{4} \sum_{a=1}^4 \sum_{b>a}^4 \abs{\vec{x}^{(a)} - \vec{x}^{(b)}}^2  ~ ,
\end{equation}
which can be easily computed as the trace of the shape tensor: $R^2(t) = \trace{\vec{G}}(t)$. The linear size is therefore determibed by pair-dispersion statistics, the study of its shape, however reveals different properties of the underlying flow. \par
The evolution of the shape tensor eigenvalues is governed by the tetrahedron's perceived velocity-gradient tensor $\vec{M}$, defined by
\begin{equation}
        \vec{M} := \vec{G}^{-1}\vec{W} - \frac{1}{3}\mathrm{tr}\qty(\vec{G}^{-1}\vec{W}) \vec{I} \qc
\end{equation}
with the identity tensor $\vec{I}$, the auxiliary tensor $W_{ij}=\sum_{a=1}^4(u^{(a)}_{i}(t) - u^c_{i}(t))(u^{(a)}_{j}(t) - u^c_{j}(t))$, where $\vec{u}^{(a)}$ is the velocity of particle $a$ and $\vec{u}^c = \sum_{a=1}^4\vec{u}^{(a)}/4$~\citep{pumirTetrahedron2013}. Similarly to the velocity gradient tensor, the tensor $\vec{M}$ can be decomposed in its symmetric and antisymmetric parts, which are called perceived rate-of-strain tensor $\vec{S} = (\vec{M}+\vec{M}^T)/2$ and and percieved perceived vorticity tensor $\boldsymbol{\Omega} = (\vec{M}- \vec{M}^T)/2 =  \sum_{k=1}^{3} \epsilon_{ijk} \omega^{l}_k(t)/2$ with the local vorticity $\boldsymbol{\omega^{l}}$. Using these symbols the short-time evolution of the shape-tensor's eigenvalues can be obtained from a Taylor-expansion~\cite{juchaTimeSymmetry2015}, which reads
\begin{equation}
    \average{g_i(t)} \approx \frac{R_0^2}{2} \qty[1 + 2 \average{s_i} t] + \mathrm{h.o.t.} \qc
  \end{equation}
  with the averaged eigenvalues $\average{s_1} \geq \average{s_2} \geq \average{s_3}$ of the perceived rate-of-strain tensor $\vec{S}$. \par

\subsection{Data analysis methodology}
\subsubsection{Identification of Lagrangian pairs and tetrahedra}
The analysis presented below rests in a crucial way on selecting particle pairs initially separated by a specific distance $r_0$, up to a tolerance $\Delta r_0$, and possibly with a separation $\vec{r}_0$ with a specific orientation, i.e., a predefined angle to the $z$-axis 
\begin{equation}
    \cos(\vartheta) = \abs{\frac{\vec{r}_0 \cdot \vec{e}_z}{r_0}} \qq*{,}
\label{eq:angle}
\end{equation}
up to a tolerance $\Delta \vartheta$. Given that an exchange between the two particles merely changes the sign of the scalar product $\vec{r}_0 \cdot \vec{e}_z$, the absolute value in \cref{eq:angle} merely ensures that  $\vartheta$ is in the range $[0,\pi/2]$. The brute-force approach to this problem would be to compute first all pairwise distances and angles $\vartheta$ to identify the pairs that fulfill the criteria. However, this approach is numerically prohibitively expensive, the more so as the number of particles increases. To overcome this difficulty, we employ a modified version of the recently proposed algorithm~\cite{gallonLagrangian2024a} using the octree-data structure~\cite{finkelQuad1974}. The algorithm is discussed in more detail in~\cref{apx:B}.
\subsubsection{Lagrangian averages and detection of spurious periodicicy effects}\label{sec:lagAVG}
\begin{figure}[t]
    \centering
    \includegraphics[width=\textwidth]{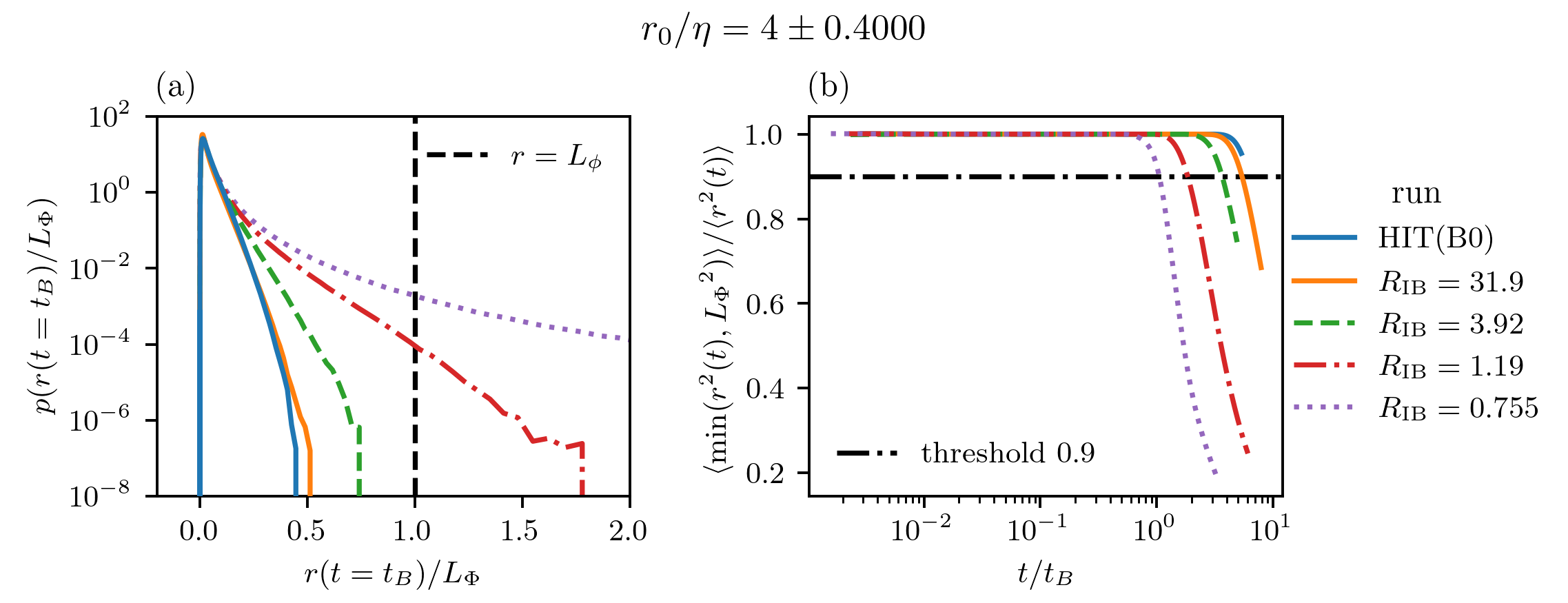}
    \caption{Removing spurious periodicity effects. Panel (a): Histograms of the absolute value of the separation vector $r(t) = \abs{\vec{r}(t)}=\abs{\vec{x}^{(2)}-\vec{x}^{(1)}}$ at time $t=t_B$ for pairs that are initially separated by $(4 \pm 0.4) \eta$. Panel (b): Data Cleaning criterion: Quotient of  the limited second moment $\langle \mathrm{min}(r^2(t),{L_B}^2)  \rangle$ and the second moment $\langle r^2(t) \rangle$ of the absolute value of the separation vector. When the curves cross the chosen threshold of $0.9$, large-scale effects contribute significantly to the second moments and the results should be interpreted carefully.} \label{fig:dataCleaning}
\end{figure}
To compute dispersion statistics from the data base of trajectories generated by the simulations described in \cref{ssc:dns}, we start by selecting a particular intermediate time $t^\prime$. We then identify particle pairs that have a relative distance of $r_0 \pm \Delta r_0$ (and in some cases orientation $\vartheta \pm \Delta \vartheta$) or regular tetrahedra with edge length $R_0 \pm \Delta R_0$ at this time $t^\prime$ using the algorithms discussed in the previous section. For the analysis presented here, we chose the tolerances $\Delta r_0$ and $\Delta \vartheta$ so, that the expected number of pairs are independent of $r$ and $\vartheta$, i.e., $\Delta r_0 \propto  r_0^{-2}$ and $\sin{\Delta\vartheta} \propto 1/\sin(\vartheta)$ and for respectively tetrahedra $\Delta R_0 \propto R_0^{-1/2}$. \par 
We then follow these particle pairs both forwards and backwards in time using the database of trajectories, to obtain estimates of the studied observables, for instance $\langle\delta_{r_0} r^2(t)\rangle$ and $\langle\delta_{r_0} r^2(-t)\rangle$. We repeat this procedure for a total of 128 equidistant points in time to accumulate for each considered initial separation a total of $\sim 10^8$ particle pairs when considering all orientations and  $\sim 10^7$ particle pairs when prescribing $\vartheta \pm \Delta \vartheta$. \par 
We recall that the simulation box is a triply periodic domain. The particle trajectories are integrated at runtime and remain as well in this domain. Therefore, when a particle is about to cross a domain border, it gets reinserted, jumping by a distance of $\pm 2 \pi$. We detect and remove these jumps whilst following the particle pairs and continue the trajectories in an enlarged domain to obtain smooth pair statistics. This, however, could lead to a contamination of the dispersion statistics, an effect we take into account in the analysis that follows. \par
Let us consider a pair of particles initially close. The two particles then separate quickly as they are advected by the turbulent flow. At a later point in time, one of the particles crosses the domain boundary and is reinserted at the other end of the domain. For the dispersion statistics, however, the  particle does not jump and remains in the continuation of the domain, as discussed before. If the two particles now happen to get close again in the simulation coordinates, they experience non-physical correlations of their velocities. This may cause problems, especially in stratified turbulent flows, where vertical separations are limited and we expect large separations in the horizontal plane~\citep{vanaartrijkSingleparticle2008}. \par
To limit the influence of such periodicity effects, we evaluate the contribution of pairs that have a separation equal to or greater than the forcing scale ${L_\Phi}$. To this end, we compare the variance of the separation, $\langle r^2(t) \rangle$, with $\langle \mathrm{min}(r^2(t),{L_\Phi}^2)  \rangle$ and $\langle r^2(t) \rangle$. 
If their quotient crosses a threshold, chosen here to be $0.9$, the highly separated pairs influence the result by more then $10\%$. In the following figures, we indicate this situation by showing the results as a dotted black lines, bringing attention to the possible bias due to spurious periodicity effects.
An example of this procedure is shown in \cref{fig:dataCleaning}.
\section{Lagrangian Multiparticle Statistics}
\label{sec:Results}

\subsection{Pair dispersion}
\label{subsec:pair_dispersion}
\begin{figure}[tb]
    \centering
\includegraphics[width=\textwidth]{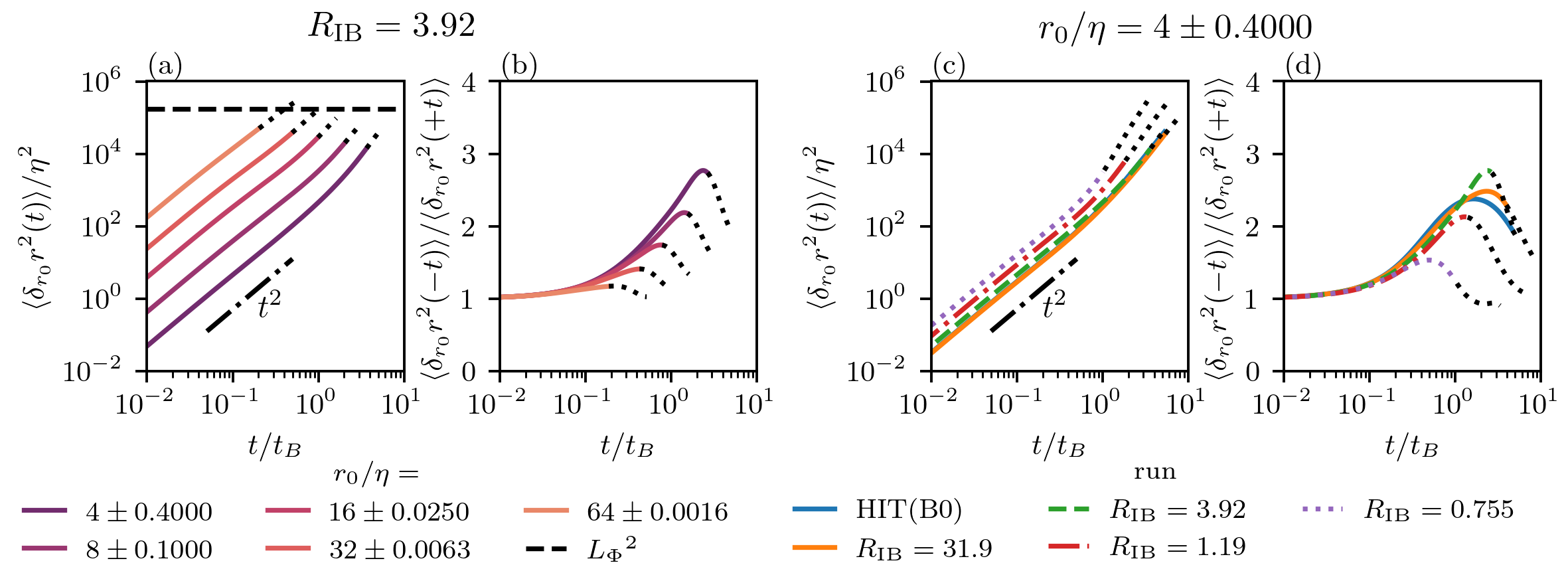}
    \caption{Panels (a,c): Mean squared change of separation $\langle \delta_{r_0}r^2(t)\rangle$ averaged over pairs with the same initial separation $r_0$. Panels (b,d): disparity between backward and forward dispersion. Panels (a,b): statistics for run B2 at different initial separations (indicated by color). Panels (c,d): Pair statics for the same initial separation $r_0/eta = 4\pm 0.4$ and different runs runs B0 - B4 (indicated by color).}\label{fig:mscosTotal}
\end{figure}
We first consider the evolution of the total mean squared change of separation $\delta_{r_0}r^2(t)$, shown in \cref{fig:mscosTotal} as a function of time $t$. As seen in the figure, $\delta_{r_0}r(t)$, grows with the expected $t^2$ scaling for short times. Furthermore, we observe that the separation grows quicker backwards than forwards, which is an indicator of a kinetic energy cascade from large to small scales~\cite{gallonLagrangian2024a}. The disparity first grows before relaxing at intermediate times. \par
\begin{figure}[tb]
    \centering
    \includegraphics[width=\textwidth]{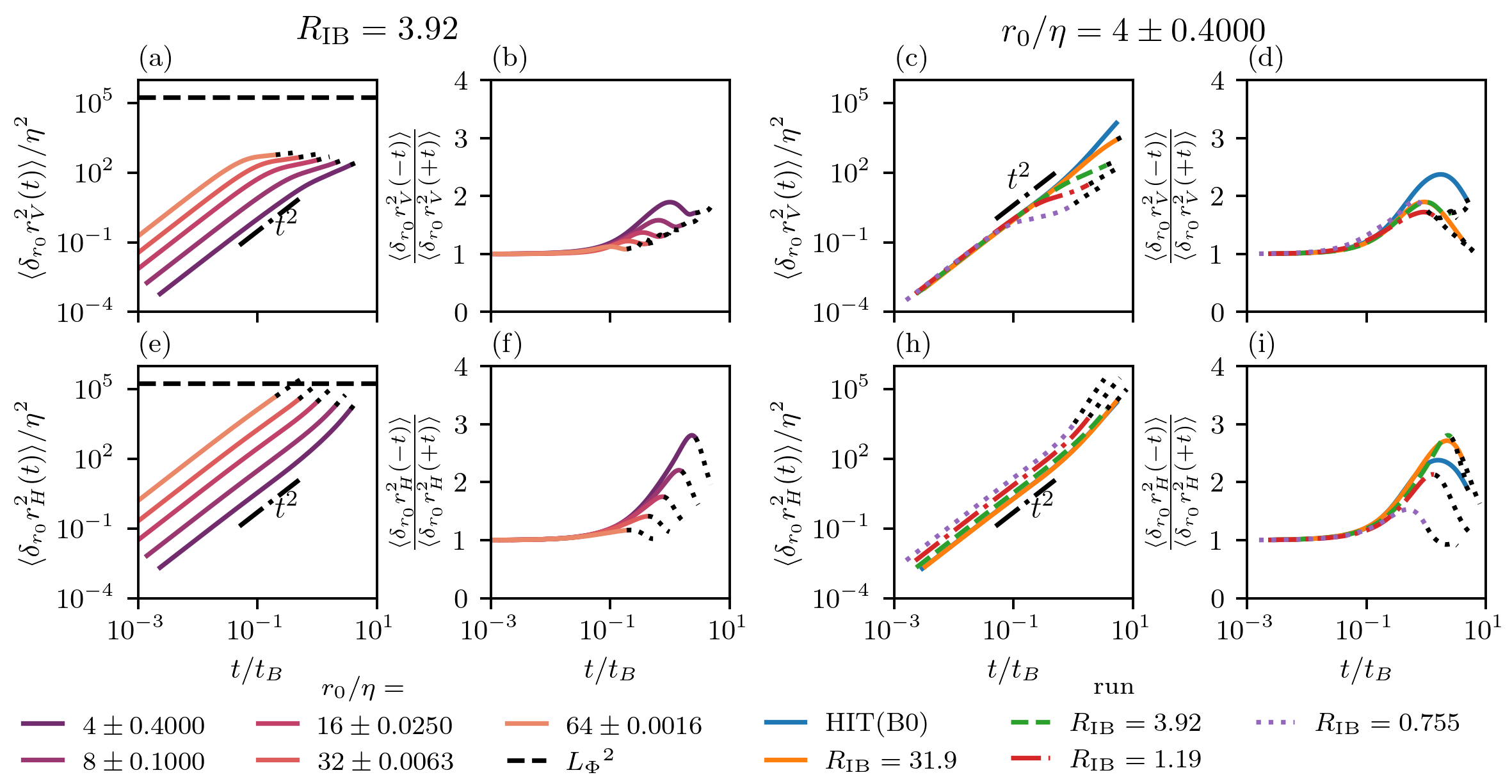}
    \caption{Anisotropy of pair dispersion First row: vertical dispersion. Panels (a,c): Vertical component of the mean squared change of separation $\langle \delta_{r_0}r_V^2(t)\rangle$, Panels (b,d): disparity between vertical backward and forward dispersion. Second row: horizontal dispersion. Panels (e,g): Horizontal component of the mean squared change of separation $\langle \delta_{r_0}r_H^2(t)\rangle$ averaged over pairs with the same initial separation $r_0$. Panels (f,i): disparity between horizontal backward and forward dispersion. Panels (a,b,e,f): statistics for run B2 at different initial separations (indicated by color). Panels (c,d,h,i): Pair statics for the same initial separation $r_0/eta = 4\pm 0.4$ and different runs runs B0 - B4 (indicated by color).}\label{fig:mscosComponents}
\end{figure}

To explore the consequences of the flows anisotropy on the dynamics of particle pairs, we split $\delta_{r_0}r^2(t)$ in its horizontal $\delta_{r_0}r_H^2(t)$ and vertical components $\delta_{r_0}r_V^2(t)$. The result is shown in \cref{fig:mscosComponents}. Similarly to the observations for purely stratified flows \cite{vanaartrijkSingleparticle2008}, we observe the vertical component to flatten at the time-scale of buoyancy fluctuations $ T_B \sim 2\pi/N$. This trend is not observed for the total mean squared change of separation, as it is dominated by its horizontal part, that does not exhibit this flattening out, behaving similarly in the rotating stratified cases as in HIT.
\subsubsection{Influence of the initial orientation}
\begin{figure}[tb]
    \centering
    \includegraphics[width=\textwidth]{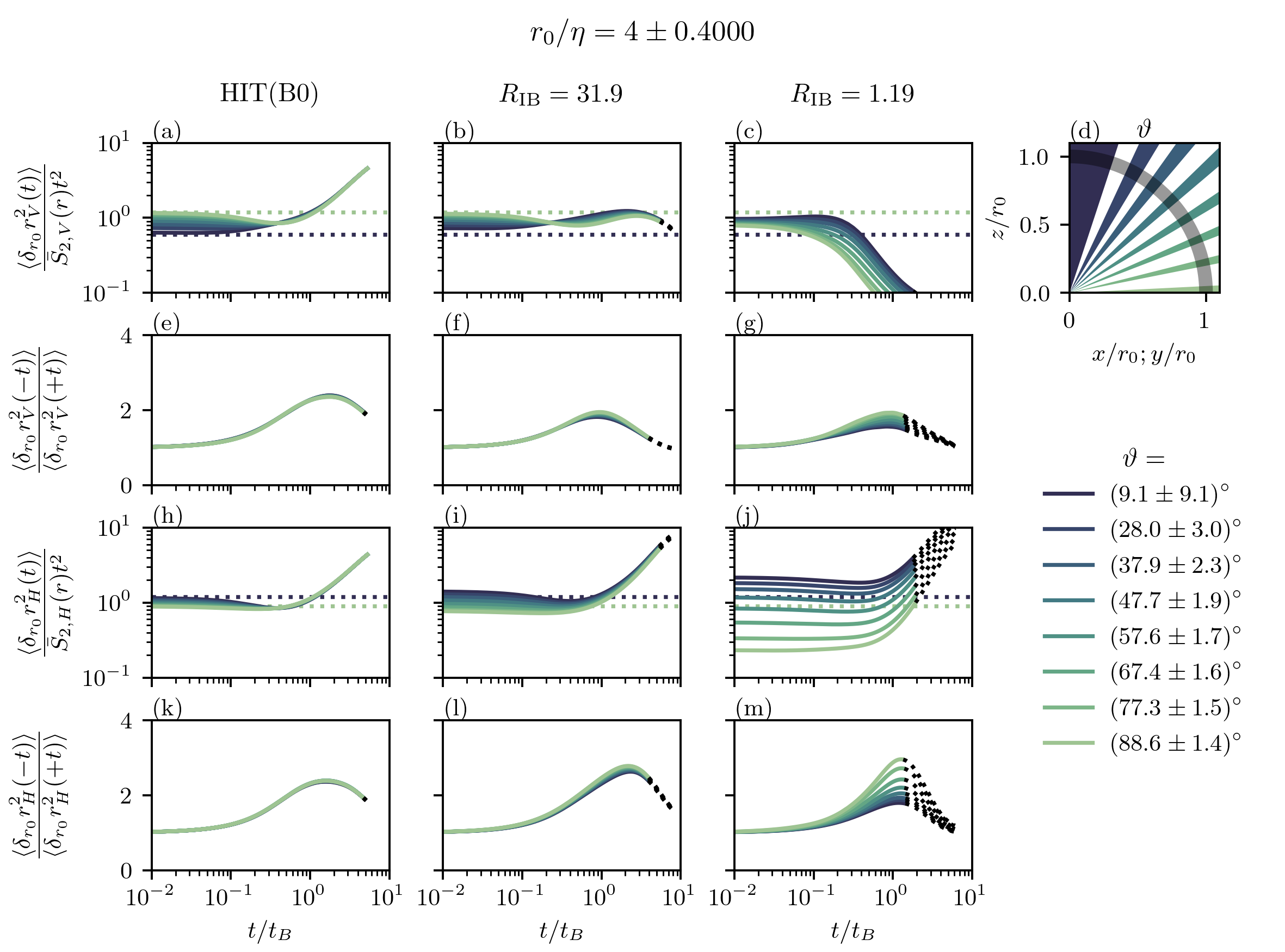}
    \caption{Influence of initial orientation on the vertical (rows 1 and 2) and horizontal (rows 3 and 4) mean squared change of separation for pairs that are initially separated by $(4 \pm 0.4)\eta$. Color indicates initial orientation $\vartheta$: from dark blue (vertical) to light green (horizontal), see panel (d). Column 1: HIT run for comparison (run B0 in \cref{tab:param}). Column 2: weakly rotating-stratified turbulence (run B1 in \cref{tab:param}), Column 3: strongly rotating-stratified turbulence (run B3 in \cref{tab:param}).}\label{fig:mscosComponentsOrientation}
\end{figure}
\begin{figure}[tb]
    \centering
    \includegraphics[width=\textwidth]{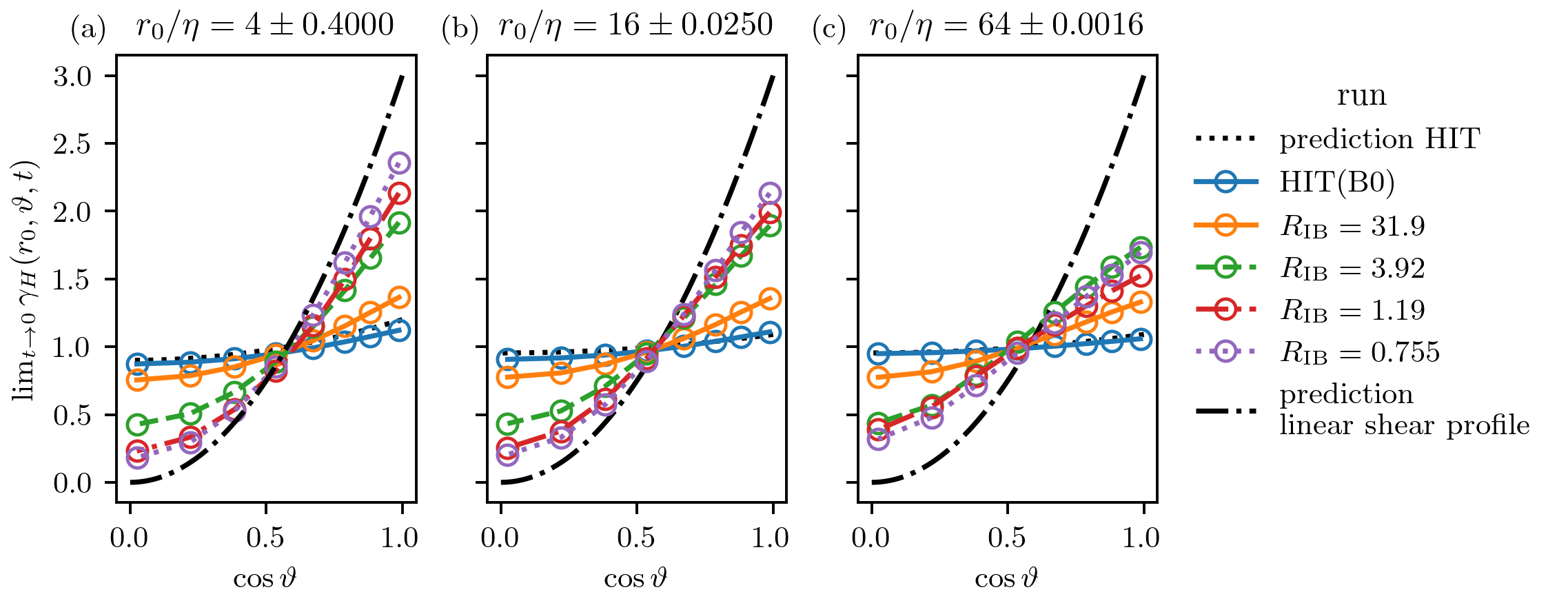}
    \caption{Influence of the initial orientation on pair dispersion. Comparison with shear flow $U_H = \alpha z $.}\label{fig:mscosComponents_shear}
\end{figure}
Rotation and stratification do not only influence the vertical and horizontal dispersion differently, but also introduce a dependency of the initial orientation on pair dispersion statistics. We parametrize the orientation by using the angle $\vartheta$ between the two particles forming the pair and the vertical axis, as defined in \cref{eq:angle}. For pairs with $\vartheta\approx 0$ the particles are aligned with the vertical axis, whereas in case $\vartheta\approx \pi/2$ the initial pair separation vector lies in the horizontal plane. In particular, we study the influence of the initial orientation on the vertical and horizontal pair dispersion. 
We remark that even in the HIT case, vertical and horizontal dispersion statistics depend on the angle $\vartheta$, as they are mixtures of transverse and longitudinal structure functions. To quantify the disparity, we define the vertical and horizontal angle-dependent structure functions and their orientation-averaged counter parts
\begin{align}
    S_{2,H}(r,\vartheta) = \average{\Delta_\vec{r} u^2} + \average{\Delta_\vec{r} v^2} &\qq{and}
    S_{2,V}(r,\vartheta) = \average{\Delta_\vec{r} w^2} \label{eq:struc_fns1}\\
    \bar{S}_{2,V}(r) = \int_{0}^{\frac{\pi}{2}} \dd{\vartheta} S_{2,V}(r,\vartheta)\sin(\vartheta) &\qq{and} \bar{S}_{2,H}(r) = \int_{0}^{\frac{\pi}{2}} \dd{\vartheta} S_{2,H}(r,\vartheta)\sin(\vartheta)
\label{eq:struc_fns2}
\end{align}
For HIT, the second-order structure functions depend solely on the longitudinal structure function $D_{LL}(r)$. Specifically, one finds for $r$ in the inertial range, i.e., $\eta \ll r \ll L$ (see \cref{apx:B})
\begin{align}
    S_{2,V}(r,\vartheta) &= \frac{4 - \cos^2(\vartheta)}{3}D_{LL}(r)  \qq{and} \bar{S}_{2,V}(r) = \frac{11}{9} D_{LL}(r)\qq*{,} \\
    S_{2,H}(r,\vartheta) &=  
    \frac{8 - \sin^2(\vartheta)}{3} D_{LL}(r) \qq{and} \bar{S}_{2,H}(r) = \frac{22}{9}D_{LL}(r) \qq*{.}
\end{align}
For $r$ in the dissipative range, i.e., $r \ll \eta$, one finds (see \cref{apx:B})
\begin{align}
    S_{2,V}(r,\vartheta) &=\qty[2 - \cos^2(\vartheta)] D_{LL}(r) \qq{and} \bar{S}_{2,V}(r) =  \frac{5}{3}D_{LL}(r) \qq*{,}\\
    S_{2,H}(r,\vartheta) &= \qty[4 - \sin^2(\vartheta)]  D_{LL}(r)  \qq{and} \bar{S}_{2,H}(r) =  \frac{10}{3}D_{LL}(r) \qq*{.}
\end{align} 
In \cref{fig:mscosComponentsOrientation}, we show the
vertical and horizontal components of the mean squared change of separation for runs B0, B1 and B3, and the initial separation $r_0/ \eta + 4 \pm 0.4$ varying the initial orientation. The vertical (panels (a-c)) and the horizontal (panels (h-j)) displacements are divided by the orientation-averaged second order structure functions ${\bar S}_H(r)$ and ${\bar S}_V(r)$, defined in \cref{eq:struc_fns1,eq:struc_fns2}, times $t^2$. As a consequence, a horizontal line corresponds to a growth $\propto t^2$. As a point of comparison, it is first useful to focus on the HIT case. With the normalization chosen, panels (a,h) of \cref{fig:mscosComponentsOrientation} show a dependence on the initial orientation of the particle difference, which is accurately captured at short times by the prediction of the angle-depended structure functions for $r_0$ in the dissipative range, see \cref{eq:stre_funct_vert,eq:stre_funct_hor,apx:B}. At longer times, the dependence on the initial orientation becomes immaterial, as shown by the overlap between the curves in panels (a,h). \par
The addition of rotation and stratification modifies the angular dependence of the separation. This is shown for runs B1 and B3, respectively in the presence of a weak (B1) and stronger (B3) stratification and rotation. As expected, the changes in the structures of the displacement compared to HIT are moderate when rotation and stratification are weak (panels (b,i)). In particular, the expectations based on HIT predictions provide a qualitative explanation of the dependence observed at short times for run B1. Interestingly, one observes an inversion at larger times in the trend for the vertical separation, whereas pairs originally vertically aligned disperse slower than those aligned horizontally, the growth parallel to gravity becoming faster for initially vertically oriented pairs at intermediate times, see \cref{fig:mscosComponentsOrientation} panel (b). We observe for the horizontal dispersion that originally vertical aligned pairs are quicker dispersed, as it was the case for HIT, but the difference between the various orientations becomes weaker at later times, i.e., for $t/t_B \gtrsim 1$, see \cref{fig:mscosComponentsOrientation} panel (i). \par 
Remarkably, at the larger value of the stratification and rotation shown in \cref{fig:mscosComponentsOrientation}, one observes a much pronounced dependence on the orientation, particularly for the dispersion in the horizontal direction, which persists at long times (panels (c,j)). In general, we remark that the angular dependence of the initial separation becomes more pronounced when buoyancy Reynolds number $\RIB$ decreases. In particular, for run B3, the dependence is very different compared to that expected from the structure function analysis, which works very well for HIT, and to some extent for moderate stratification. This is a manifestation of the qualitative change of behavior of the flow, which becomes dominated by waves, not by eddies, when $\RIB$ becomes of order $1$. \par
In general, the differences between forward and backward dispersion are insensitive to the initial orientation of the particle pairs for moderate stratification and rotation. This is illustrated by panels (f,l), which show at most a very weak difference between the curves corresponding to the various orientations, the dispersion backwards in time being generally faster than forward in time, as already observed for HIT (panels (e,k)). A completely different picture is observed when the stratification is enhanced, see panels (g) and especially (j). Overall, the symmetry breaking between backwards and forwards dispersion is much stronger for particles initially in the horizontal direction. \par

A remarkable feature revealed by the conditioning dispersion on the angle between the vertical and the separation between the two particles is the appearance of the very strong dependence of $\langle \delta_{r_0} r^2_H(t) \rangle$, the separation in the horizontal plane, with the angle, $\vartheta$, when $\RIB$ becomes of order $1$, see \cref{fig:mscosComponentsOrientation} panel (j). The very systematic increase of $\langle \delta_{r_0} r^2_H(t) \rangle$ with $\vartheta$ could be attributed to a dominant shear, which would separate most efficiently particles the furthest apart in the vertical direction. This possibility has been proposed several times in the literature, in turbulent flows with a strong stratification see e.g.,~\cite{lillyStratified1983,billantExperimental2000,billantThreedimensional2000,rileyStratified2008}. To test more quantitatively the hypothesis that the flow is dominated by a shear when stratification increases, we consider the ratio 
\begin{equation}
    \gamma_H(r_0,\vartheta,t)= \frac{\langle \delta_{r_0}r_H^2(t)\rangle}{\bar{S}_{2,H}(r_0)}
\end{equation}
that is shown in \cref{fig:mscosComponentsOrientation} panels (i-j) as a function of time, and take the limit $t \to 0$. We show this limit in \cref{fig:mscosComponents_shear} as a function of $\cos\vartheta$, for three initial distances: $r_0 = (4 \pm 0.4) \eta$ (panel (a)) and $r_0 = (16 \pm 0.025) \eta$ (panel (b)), for the 5 runs at high resolution, B0-B4. The prediction for HIT flows is shown as a black dotted lines, and describes well the data for HIT flows (run B0). We compare the measured ratio to the case of a linear shear profile with $U_H(z) = \alpha z$ with a constant $\alpha$. In this case, we find $S_{2,H}(r,\vartheta)/\bar{S}_{2,H}(r)=\alpha^2 r^2 \cos^2(\vartheta)/ \qty(\alpha^2 r^2/3) =3 \cos^2(\vartheta)$, which is shown in black dashed-dotted lines. The clear trend observed, for both values of $r_0$, is that the curves continuously evolve from the HIT limit (for B0) towards the case corresponding to the pure shear.  The effect, however, becomes weaker when $r_0$ increases. \par 
Our numerical work therefore confirms the prevalence of a shear in turbulent flows with a large stratification~\cite{lillyStratified1983,billantExperimental2000,billantThreedimensional2000,rileyStratified2008} and extends the result to flows with weak rotation. Such observation extends all the way down to $\RIB \lesssim 1$. An interesting question is whether this conclusion still applies at even larger values of the stratification, when the flow becomes essentially laminar.
As far as oceanic flows are concerned, the values of the Reynolds numbers are as large as $10^9$~\cite{thorpeIntroduction2007}, with values of $\RIB$ of the order of $100$. How much of the picture drawn in \cref{fig:mscosComponentsOrientation} applies in this case remains to be understood, although it is safe to think that as the system attains values of $\RIB$ larger than 10 and closer to 100, it enters a regime where the dynamics reproduced by DNS are more realistic than at smaller values of $\RIB$.
\subsubsection{Non-Gaussian statics of pair dispersion}
The study of the mean squared separation provides good estimates on the separation dynamics taken by most pairs. However, the intermittent character of turbulent flows can separate few pairs much quicker than the average. To study such non-Gaussian effects in pair dispersion, we consider the kurtosis of the separation distance $r(t) = |\vec{r}(t)|$ of particle pairs initially separated by $\vec{r}_0$~\cite{biferaleLagrangian2005}
\begin{equation}
    K_r(\vec{r}_0,t) := \frac{\left\langle \left(r(t) - \bar{r}(t)\right)^4 \right\rangle}{\left\langle (r(t) - \bar{r}(t))^2 \right\rangle^2} \qq{where}  \bar{r}(t) = \average{r(t)} \qq*{.}
\end{equation}

We remark that other observables can be used as well to study this intermittent property of pair dispersion, see for instance~\cite{buariaCharacteristics2015,bitaneGeometry2013}. \par
\begin{figure}[t]
    \centering
    \includegraphics[width=\textwidth]{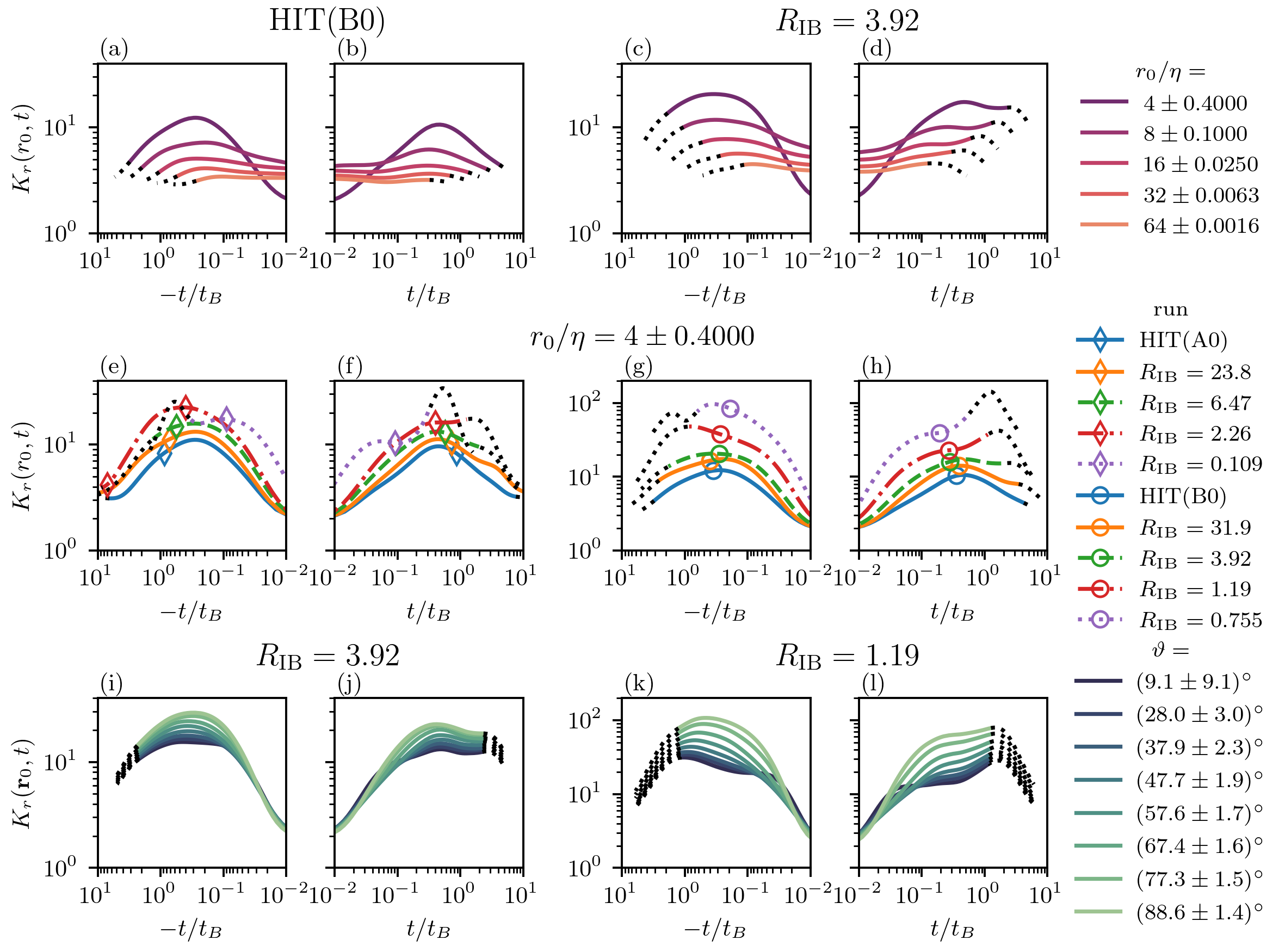}
    \caption{Kurtosis of pair separation forwards in time panels (b,d,f,h,j,l) and backwards in time panels (a,c,e,g,i,k). The first row shows the dependence on the initial orientation $r_0$ for HIT panels (a,b) and moderate rotation and stratification panels(c,d). The second row shows the dependence on $R_\mathrm{IB}$, note the rescaled $y$ axis in panels (g,h). The third row shows the dependence on initial separation for low $R_\mathrm{IB}$ (panels (i,j)) and strong (panels(k,l)) rotation and stratification. We remark that the lines are continued in dotted lines, when more than $10\%$ of the kurtosis is caused by potentially spurious large-scale effects. Panels (a-h) are averaged over $\sim 10^8$ particle pairs per initial separation or run. Panels (i-l) are averaged over $\sim 10^7$ particle pairs per orientation
}\label{fig:kurtosis}
\end{figure}

\Cref{fig:kurtosis} shows the time evolution of $K_{r}(\vec{r}_0,t)$ 
and its dependence on the initial separation (panels (a-d)), on $\RIB$ (panels (i-l)) and on the initial orientation $\vartheta$ (panels (e-h)).
Overall, $K_{r}(r_0,t)$ reaches larger values when the separation $r_0$ is smaller. This is the case for both HIT flows (panels (a,b)) and for flows in the presence of rotation and stratification (panels (c,d)). Similar trends are observed for the evolution backwards (panels (a,c)) and forwards (panels (b,d)) in time, the former leading to slightly larger values of $K_{r}$. Comparing the cases HIT and moderately strong rotation and stratification suggests that the kurtosis in fact increases with increasing stratification and rotation, reaching values as high as $\approx 20$ for $\RIB = 3.92$. This is consistent with panels (e-h), which show the dependence on $\RIB$ at the initial separation $r_0 = 4 \pm 0.4 \eta$, confirming the suggestion of panels (a-d) that the kurtosis becomes larger when stratification and rotation increase. Furthermore, the comparison of panels (e,f) to panels (g,h) in \cref{fig:kurtosis} shows that the pair kurtosis depended sensitively on the Reynolds number $\mathrm{Re}$: The runs B0-B4 that are run at higher resolution (shown in panels (g,h)) reach significantly higher values than the comparable simulations at the lower resolution shown in panels (e,f). In fact, at the strongest values of the stratification and rotation, corresponding to the values of $\RIB < 1$, $K_{r}$ reaches extremely large values of order $\sim 100$. An example of the probability density function with very large tails is shown in \cref{fig:dataCleaning}. We recall that the contribution of pairs separated by a distance larger than the forcing scale is less than $10\%$ of the different curves shown in full in \cref{fig:kurtosis} (see \cref{sec:lagAVG}). Panels (i-l) show the dependence on the angle between the initial pair separation and the vertical axis, for two runs with a relatively strong stratification: $\RIB = 3.92$ (panels (i,j)) and $\RIB = 1.92$ (panels (k,l)). These panels reveal that pairs originally horizontally oriented lead to the largest values of the kurtosis $K_{r}$. For an HIT flow, the effect is absent and we observe in our data a maximum relative deviation of $2.5 \%$ from the orientation average. For the rotating stratified flows, we observe that the difference as a function of the angle is strongly enhanced when the stratification becomes large, i.e., $\RIB \rightarrow 1$.

\subsection{Tetrahedra dispersion}
This section is devoted to the statistical properties of the deformation of fluid volumes at moderately long times. Indeed, having our simulations been performed at high resolution (on grids of $512^3$ and $1024^3$ points) and with $\sim 10^6$ Lagrangian particles, the runs analyzed here are not sufficiently long to reliably extract the properties at long times, contrary to what was done in~\cite{polancoMultiparticle2023}, based on $256^3$ simulations of rotating flows (with no stratification) and $\sim 10^5$ Lagragian tracers. Instead, we focus on dispersion at intermediate times, and in particular, on the forward-backward time asymmetry reflecting the fundamental irreversibility of the flow and the alignment with the preferential axis. 
We recall that, consistent with the notation introduced in \cref{subsubsec:volume_def}, $R_0$ denotes the length of the edge of initially regular tetrahedra.
\subsubsection{Forward-backward time asymmetry at short time}

\begin{figure}[t]
    \centering
\includegraphics[width=\textwidth]{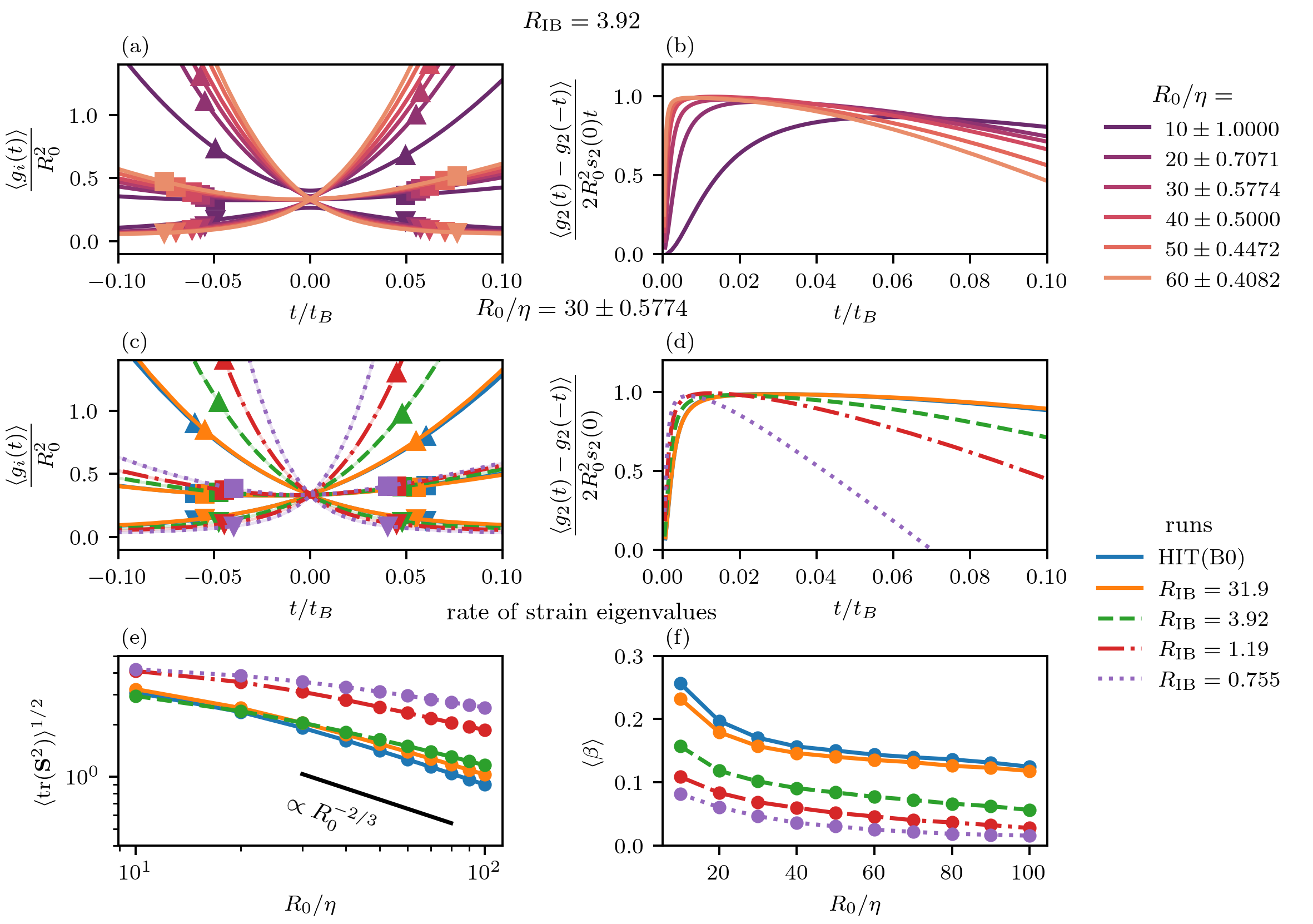}
\caption{Irreversibility of the shape tensors eigenvalues for initially regular tetrahedra. Panels (a,c) show the averaged eigenvalues $g_i(t)/R_0^2$ as a function of $t$. Panels (b,d) show the time-reversal asymmetry of the intermediate eigenvalue as a function of $t$. Different colors indicated different initial edge length for panels (a,b) and different $R_\mathrm{IB}$ for panels (c,d). Panels (e,f) show properties of the perceived velocity gradient tensor}\label{fig:shapeTensorAsym}
\end{figure}
The results presented in \cref{fig:shapeTensorAsym} focus on the time asymmetry in the deformation of tetrahedra and extend those obtained for HIT flows~\cite{juchaTimereversalsymmetry2014} to turbulent flows in the presence of rotation and stratification. \Cref{fig:shapeTensorAsym} panels (a,c) show the eigenvalues of the moment of inertia tensors, $g_1(t)$ (topmost line with up-pointing triangles) , $g_2$  (intermediate line with squares) and $g_3$ (lowest line with up-pointing triangles), with varying the initial edge length for run B2 in panel (a) and for the fixed initial edge length $R_0 = 30 \eta$ and for the 5 runs at high resolution B0-B4, including the HIT case (B0). Here, we normalized time by $t_B$ defined in \cref{eq:tBatchelor}. The influence of stratification and rotation is quite strong on $g_1$, and weaker on $g_2$ and $g_3$. This suggests that the deformation of the tetrahedra occurs faster when stratification and rotation increase. \par
As shown in~\cite{juchaTimereversalsymmetry2014}, the evolution of $g_2(t)$, is sensitive to the intermediate eigenvalue of the rate-of-strain tensor. Specifically, by ordering the eigenvalues $s_i$ of the course grained rate-of-strain in decreasing order ($s_1 \ge s_2 \ge s_3$), one observes that $s_2$ is predominantly positive~\cite{betchovInequality1956, pumirTetrahedron2013}. This, in turn, implies that $g_2$ increases linearly in time close to $t = 0$: $g_2(t) - g_2(-t) \approx 2 R_0^2 \langle s_2 \rangle t$ and therefore leads to a clear manifestation of irreversibility of the flow. The ratio $(g_2(t) - g_2(-t) )/ ( 2 R_0^2 \langle s_2\rangle)$, shown in \cref{fig:shapeTensorAsym} panels (b,d), reaches a value close to $1$ after a very short time. We notice that the value $0$ at $t/t_B \to 0$ is a consequence of the small discrepancy between the almost regular tetrahedra identified by our algorithm (allowing for deviations up to $\Delta R_0$) and perfectly regular tetrahedra. This ratio remains $\sim 1$ over a duration $~\sim 0.1 \, t_B$ for HIT and weak rotation and stratification. When decreasing $\RIB$, i.e., increasing the strength of rotation and stratification, the duration, expressed in units of $t_B$, where the discussed ratio is close to $1$ decreases. Nonetheless, even at $\RIB = 1.19$, the curve shown in \cref{fig:shapeTensorAsym} panel (d) exhibits an extended range up to $\sim 0.05 \, t_B$ where the discussed ratio is $\sim 1$. Consistent with several other observations~\cite{buariaSingleparticle2020,gallonLagrangian2024a}, a transition to a wave dominated regime occurs for $\RIB \sim 1$. \par
In view of the importance of the rate-of-strain tensor in the analysis above, \cref{fig:shapeTensorAsym} panels (e,f) illustrate some features of the perceived rate-of-strain tensor's eigenvalues  as a function of $R_0$. For HIT, the rate-of-strain tensor's eigenvalues $\langle s_i \rangle $ and consequently also $\langle \trace{\vec{S}^2}\rangle^{1/2}$, vary as $R_0^{-2/3}$ (see \cref{fig:shapeTensorAsym} panel (e) ) which is consistent with classical phenomenological considerations~\cite{pumirTetrahedron2013}. The $R_0^{-2/3}$ dependence captures qualitatively the dependence of $\langle \trace{\vec{S}^2}\rangle^{1/2}$ as a function of $R_0$ for $\RIB \sim 30$, and to a weaker extent, for $\RIB \sim 4$. Very significant deviations with respect of this dependence on $r_0$ are clearly visible for $\RIB \lesssim 1$. Then, $\langle \trace{\vec{S}^2}\rangle^{1/2}$ shows a slower decay with $R_0$, which is consistent with the observed faster deformation of the tetrahedra. \par
 It is also of intrinsic interest to compare the contributions of the individual eigenvalues of perceived rate-of-strain, in the presence of stratification and rotation. In view of the convention $s_1 \ge s_2 \ge s_3$, and of the incompressibility condition: $\langle s_1 + s_2 + s_3\rangle = 0$, $\langle s_1 \rangle$ is always positive. Moreover, the intermediate eigenvalue, which is related to vortex stretching~\cite{betchovInequality1956,ashurstAlignment1987,pumirTetrahedron2013}, is on average slightly positive. To measure the relative importance of this effect, \cref{fig:shapeTensorAsym} panel (f) shows the quantity $\beta = \sqrt{6}\langle s_2 \rangle/\langle \trace{\vec{S}^2}\rangle^{1/2}$~\cite{ashurstAlignment1987}. The ratio shown in \cref{fig:shapeTensorAsym} panel (f) is also always positive, which shows that stratification and rotation do not affect the condition $\langle s_2 \rangle > 0$, a fundamental property of turbulent flows~\cite{betchovInequality1956,ashurstAlignment1987,pumirTetrahedron2013}. In general, however, the value $\beta$ decreases when stratification and rotation increase (when $\RIB $ decreases). Furthermore, \cref{fig:shapeTensorAsym} panel (f) shows that when $R_0$ grows the ratio $\beta \to 0$ and consequently $\langle s_2 \rangle \to 0$, which is expected assuming the velocities decorrelate on large scales.
\subsubsection{Tetrahedra orientation alignment}
\begin{figure}[t]
    \centering
    \includegraphics[width=\textwidth]{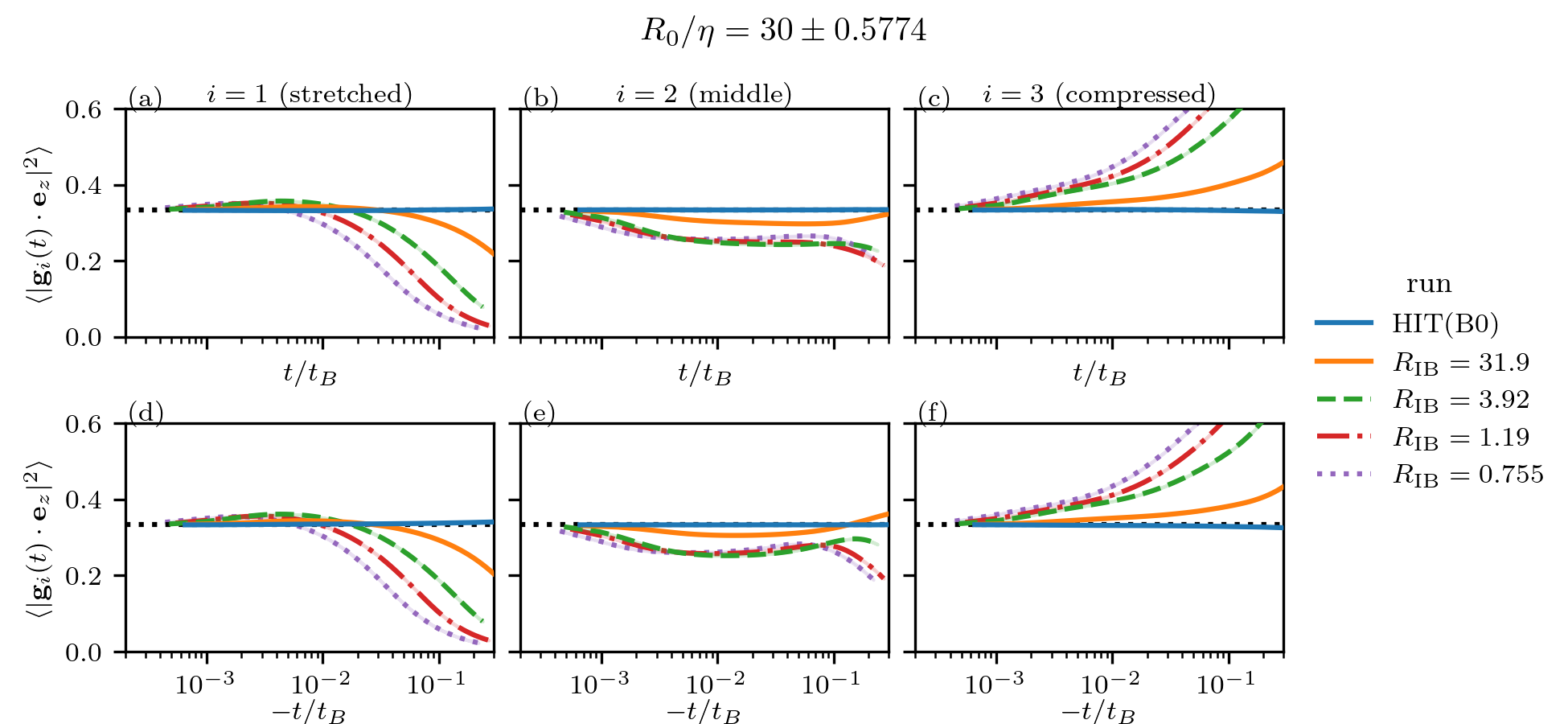}
    \caption{Preferential alignment of the perceived rate-of-strain strain tensor eigenvectors (panels (a-c)) and local vorticity (panel (d)). The reference value for random orientation is $1/3$ and indicated by a black dotted line.}
\label{fig:tetrads_orient}
\end{figure}
As a result of the flows anisotropy caused by rotation and stratification, the deformation of the tetrahedra is not isotropic. A similar feature was observed in the case of rotating turbulence~\cite{polancoMultiparticle2023}. In this subsection, we study how tetrahedra align with respect to the preferential direction of the flow, i.e., the vertical direction. \Cref{fig:tetrads_orient} shows the alignment of the eigenvectors of the shape tensor, $\mathbf{g}$ with $\mathbf{e}_z$, forwards (panels (a-c)) and backwards (panels (d-f)) in time. \par
At long times, \cref{fig:tetrads_orient} panels (c,f) show a clear preferential alignment of the most compressed eigendirection corresponding to the weakest eigenvalue $\mathbf{g}_3$ with $\mathbf{e}_z$. Simultaneously, the eigenvector corresponding to the strongest eigenvalue, $g_1$, becomes preferentially perpendicular to $\mathbf{e}_z$, see \cref{fig:tetrads_orient} panels (a,d). The effect, which is absent for HIT, becomes stronger as stratification and rotation increase, or equivalently, as $\RIB$ diminishes. In comparison, the eigenvector corresponding to the intermediate eigenvalue shows only a mild tendency to become perpendicular to $\mathbf{e}_z$ at long times, see \cref{fig:tetrads_orient} panels (b,e). \par
The main tendency is therefore that tetrahedra are mostly elongated in the horizontal plane, i.e., perpendicular to the axis of gravity and rotation, $\mathbf{e}_z$. This result is coherent with the observation in \cref{subsec:pair_dispersion} that the dispersion in the horizontal direction is much larger than the vertical one (see \cref{fig:mscosComponents}). Interestingly, our result should be contrasted with the conclusion of~\cite{polancoMultiparticle2023} that in the presence of solely rotation, tetrahedra tend to align with the axis of rotation, $\mathbf{e}_z$, the difference in the two cases being thus attributable to stratification, which is prevalent in our case ($N/f = 5$). \par
At short times, we observe, more surprisingly, a weak tendency of the intermediate eigenvector $\vec{g_2}$ to be anti-aligned with $\vec{e}_z$, see \cref{fig:tetrads_orient} panels (b,e). As a consequence, the two other eigenvectors of $\vec{G}$ are mildly aligned with $\vec{e}_z$ at short times. 

This property is consistent with the observed alignment of the vorticity and intermediate strain eigenvectors with the directions of the flow, shown in \cref{fig:aignt_strain} in \cref{AppendixB}. This property should be contrasted with the alignments reported in the case of purely rotating flows, which lead instead to an alignment of $\vec{s}_2$ with $\vec{e}_z$~\cite{nasoMultiscale2019}, and to an opposite trend in the alignment of the eigenvectors of $\vec{G}$ at short times~\cite{polancoMultiparticle2023}.
\section{Conclusion}
\label{sec:conclusion}

The presence of rotation and stratification modifies the properties of turbulent flows in many setups of geophysical significance. Our study focuses in particular on the Lagrangian properties of fluid turbulence, an aspect that has proven very fruitful in the study of homogeneous isotropic turbulence, see e.g.~\cite{toschiLagrangian2009}, but less utilized to investigate flows of geophysical relevance. Specifically, we extend the work of \cite{buariaSingleparticle2020}, devoted to single particle dispersion, to the motion of more than one particle in flows with rotation and stratification. By considering the role of stratification and rotation simultaneously, our results also extend those for flows with pure stratification~\cite{vanaartrijkSingleparticle2008} and for flows with pure rotation~\cite{polancoMultiparticle2023}. Our study is based on direct numerical simulations of the Boussinesq equations at moderate/high resolution, where we also followed the motion of up to  $N_p \sim 5 \cdot 10^6$. We then used techniques based on the octree data structure~\cite{finkelQuad1974} to identify pairs of particles separated by a given distance $r_0$, and sets of four equidistant particles, forming regular tetrahedra with an edge length $R_0$. While a direct search is prohibitive, as it involves at least $\sim N_p^2$ operations, the octree-based technique provides an efficient way to identify such particle sets, reducing computational costs by several orders of magnitude compared to brute force methods. With the methods, we could also study properties, such as the irreversibility in the evolution of sets of particles, by directly comparing the forward and backward evolution~\cite{juchaTimereversalsymmetry2014,gallonLagrangian2024a}.\par
At the fixed ratio between stratification and rotation chosen here, $N/f =5$ (compatible with the values measured in the Southern Ocean~\cite{garabatoWidespread2004,nikurashinRoutes2013}), we observe that increasing the stratification reduces the growth of the particle separation in the vertical direction, $\langle \delta_{r_0} r^2_V(t) \rangle$ at times $t \gtrsim 1/N$, whereas the horizontal separation, $\langle \delta_{r_0} r^2_H(t) \rangle$ grows qualitatively similarly to the case of an HIT flow. We observe as well that the ratio between backward and forward dispersion, which provides a measure of the time irreversibility of the flow, becomes closer to $1$ as stratification strengthens. \par
We also conditioned our statistics on the angle between the pair separation and the vertical direction (given by gravity and the axis of rotation), whilst splitting the mean squared change of separation $\langle \delta_{r_0} r^2(t) \rangle$ in its horizontal $\langle \delta_{r_0} r^2_H(t) \rangle$ and vertical $\langle \delta_{r_0} r^2_V(t) \rangle$ contributions. Remarkably, the effect of stratification and rotation on the pair dispersion statistics becomes particularly strong for the horizontal dispersion and when the buoyancy Reynolds number $\RIB \sim 1$. 
Our numerical results reveal that the horizontal displacement increases, as the initial separation between the particles becomes more aligned with the vertical direction. Interestingly, our work provides a clear confirmation that the motion is dominated by a vertical shear of the horizontal velocity~\cite{lillyStratified1983,billantExperimental2000,billantThreedimensional2000,rileyStratified2008,Marino_2014}. The comparably small amount of rotation ($N/f = 5$) should not affect this result qualitatively. The fluid motion becomes closer to a shear flow, the more so as the stratification increases, at least at scales $r_0 < L_B$ and down to values of $\RIB \lesssim 1$, i.e., in situations where the fluctuations of the velocity gradient are comparable to the Brunt-V\"ais\"ala frequency. How strongly these conclusions extend to the range $100 \le R_{\mathrm{IB}} \le 1000$ 
of $R_{\mathrm{IB}}$, corresponding to measurements in the ocean~\cite{moumEnergycontaining1996} or to similar ${R}_\mathrm{IB}$, but with both stronger turbulence (larger Reynolds number $\mathrm{Re}$) and stronger stratification and rotation (lower Froude $\mathrm{Fr}$ and Rossby $\mathrm{Ro}$ numbers), remains to be seen. \par 
Unexpectedly, we also observe that the higher moments of particle separation, in particular the normalized fourth order central moment of the separation  (the kurtosis $K_r$) is an increasing function of stratification and rotation. This is surprising, as when stratification increases the turbulent fluctuations are expected to be weaker, although recent works have shown that stratified turbulent flows can develop strong enhancements of vertical velocity and temperature, at least for certain level of stratification, leading to non-Gaussian statistics \cite{feraco_18,Pouquet_2019,feraco_21,marino_22}. Instead, we observe that the flow organizes itself, when the stratification is large, in horizontal layers with a strong shear, which can lead to very large and intense separation between two particles. We notice, however, that in our study, the values of $\RIB$ was always $\RIB \gtrsim 0.5$; it would be interesting to extend this to values with even more stratification, hence with $\RIB$ even smaller. \par

 Finally, we also studied the deformation of originally regularly shaped fluid volumes by considering tetrahedra defined by groups of four particles. At small times, this provides a way to measure the irreversibility of the flow, which has been tested for homogenous isotropic flows~\cite{juchaTimereversalsymmetry2014}. Our results show that the presence of rotation and stratification does not affect the qualitative picture, and that the manifestation of irreversibility in the deformation of tetrahedra remains strong, down to the smallest values of $\RIB$ considered here. Moreover, we observed how tetrahedra align with the flow's preferential direction, i.e., the vertical axis. 
For short times, the intermediate mode is weakly dealigned with the vertical axis, which we relate to a dealignment of the intermediate perceived rate-of-strain eigenvector. 
For longer times, we observe a pronounced dealignment of the tetrahedra's direction of greatest extend and a alignment of the direction of weakest extend with the vertical axis, or in other words, horizontally aligned needle-like tetrahedra. This observation is consistent with the observations that the pairs separate mostly in the horizontal direction. \par
To conclude, we stress that the dispersion of two or more particles, studied in one particular type of flow, is very sensitive to the detailed properties of turbulence: Our results show clear quantitative differences with the paradigmatic case of HIT. This calls for some care when modeling dispersion using results obtained from HIT. Further work appears as necessary to understand the respective role of stratification and rotation in other flows of geophysical interest.
\section{Acknowledgements}
\noindent 
S.G. acknowledges the “Fond Recherche” of \'Ecole Normale Sup\'erieure de Lyon
for financial support. R.M. and F.F. acknowledge support from the project ``EVENTFUL'' (ANR-20-CE30-0011), funded by the French ``Agence Nationale de la Recherche'' - ANR through the program AAPG-2020. A.P. was supported by the ANR project TILT (ANR-20-CE30-0035). The computing resources utilized in this work were provided by PMCS2I at the \'Ecole Centrale de Lyon and PSMN at \'Ecole Normale Sup\'erieure de Lyon.
\bibliographystyle{ieeetr}

\begin{thebibliography}{10}

\bibitem{pedloskyGeophysical1979}
J.~Pedlosky, {\em Geophysical {{Fluid Dynamics}}}.
\newblock Berlin: Springer, 1979.

\bibitem{weissTransport2008}
J.~B. Weiss and A.~Provenzale, eds., {\em Transport and {{Mixing}} in
  {{Geophysical Flows}}: {{Creators}} of {{Modern Physics}}}.
\newblock Berlin, Heidelberg: Springer, 2008.

\bibitem{guastoFluid2012}
J.~S. Guasto, R.~Rusconi, and R.~Stocker, ``Fluid {{Mechanics}} of {{Planktonic
  Microorganisms}},'' {\em Annual Review of Fluid Mechanics}, vol.~44,
  no.~Volume 44, 2012, pp.~373--400, 2012.

\bibitem{diazSeasonal2021}
B.~P. Diaz, B.~Knowles, C.~T. Johns, C.~P. Laber, K.~G.~V. Bondoc, L.~Haramaty,
  F.~Natale, E.~L. Harvey, S.~J. Kramer, L.~M. Bola{\~n}os, D.~P. Lowenstein,
  H.~F. Fredricks, J.~Graff, T.~K. Westberry, K.~D.~A. Mojica,
  N.~Ha{\"e}ntjens, N.~Baetge, P.~Gaube, E.~Boss, C.~A. Carlson, M.~J.
  Behrenfeld, B.~A.~S. Van~Mooy, and K.~D. Bidle, ``Seasonal mixed layer depth
  shapes phytoplankton physiology, viral production, and accumulation in the
  {{North Atlantic}},'' {\em Nature Communications}, vol.~12, no.~1, p.~6634,
  2021.

\bibitem{salleeSummertime2021}
J.-B. Sall{\'e}e, V.~Pellichero, C.~Akhoudas, E.~Pauthenet, L.~Vignes,
  S.~Schmidtko, A.~C. Naveira~Garabato, P.~Sutherland, and M.~Kuusela,
  ``Summertime increases in upper-ocean stratification and mixed-layer depth,''
  {\em Nature}, vol.~591, no.~7851, pp.~592--598, 2021.

\bibitem{wihsgottObservations2019}
J.~U. Wihsgott, J.~Sharples, J.~E. Hopkins, E.~M.~S. Woodward, T.~Hull,
  N.~Greenwood, and D.~B. Sivyer, ``Observations of vertical mixing in autumn
  and its effect on the autumn phytoplankton bloom,'' {\em Progress in
  Oceanography}, vol.~177, p.~102059, 2019.

\bibitem{thorpeIntroduction2007}
S.~A. Thorpe, {\em An {{Introduction}} to {{Ocean Turbulence}}}.
\newblock Cambridge: Cambridge University Press, 2007.

\bibitem{marino_15}
R.~Marino, A.~Pouquet, and D.~Rosenberg, ``Resolving the paradox of oceanic
  large-scale balance and small-scale mixing,'' {\em Phys. Rev. Lett.},
  vol.~114, p.~114504, 2015.

\bibitem{rosenberg_15}
D.~Rosenberg, A.~Pouquet, R.~Marino, and P.~Mininni, ``Evidence for
  {B}olgiano-{O}bukhov scaling in rotating stratified turbulence using
  high-resolution direct numerical simulations,'' {\em Phys. Fluids}, vol.~27,
  p.~055105, 2015.

\bibitem{moninStatistical1975}
A.~S. Monin and A.~M. Yaglom, {\em Statistical {{Fluid Mechanics}}, {{Volume
  II}}: {{Mechanics}} of {{Turbulence}}}, vol.~2.
\newblock Cambridge: MIT Press, 1975.

\bibitem{sawfordTurbulent2001}
B.~L. Sawford, ``Turbulent {{Relative Dispersion}},'' {\em Annual Review of
  Fluid Mechanics}, vol.~33, no.~Volume 33, 2001, pp.~289--317, 2001.

\bibitem{yeungLagrangian2002}
P.~K. Yeung, ``Lagrangian {{Investigations}} of {{Turbulence}},'' {\em Annual
  Review of Fluid Mechanics}, vol.~34, no.~Volume 34, 2002, pp.~115--142, 2002.

\bibitem{toschiLagrangian2009}
F.~Toschi and E.~Bodenschatz, ``Lagrangian {{Properties}} of {{Particles}} in
  {{Turbulence}},'' {\em Annual Review of Fluid Mechanics}, vol.~41, no.~1,
  pp.~375--404, 2009.

\bibitem{frischTurbulence1995}
U.~Frisch, {\em Turbulence: {{The}} Legacy of {{A}}. {{N}}. {{Kolmogorov}}}.
\newblock Cambridge University Press, 1995.

\bibitem{Marino_2013}
R.~Marino, P.~D. Mininni, D.~Rosenberg, and A.~Pouquet, ``Inverse cascades in
  rotating stratified turbulence: Fast growth of large scales,'' {\em
  Europhysics Letters}, vol.~102, p.~44006, jun 2013.

\bibitem{garabatoWidespread2004}
A.~C. Naveira~Garabato, K.~L. Polzin, B.~A. King, K.~J. Heywood, and
  M.~Visbeck, ``Widespread {{Intense Turbulent Mixing}} in the {{Southern
  Ocean}},'' {\em Science}, vol.~303, no.~5655, pp.~210--213, 2004.

\bibitem{nikurashinRoutes2013}
M.~Nikurashin, G.~K. Vallis, and A.~Adcroft, ``Routes to energy dissipation for
  geostrophic flows in the {{Southern Ocean}},'' {\em Nature Geoscience},
  vol.~6, no.~1, pp.~48--51, 2013.

\bibitem{buariaSingleparticle2020}
D.~Buaria, A.~Pumir, F.~Feraco, R.~Marino, A.~Pouquet, D.~Rosenberg, and
  L.~Primavera, ``Single-particle {{Lagrangian}} statistics from direct
  numerical simulations of rotating-stratified turbulence,'' {\em Physical
  Review Fluids}, vol.~5, no.~6, p.~064801, 2020.

\bibitem{nasoMultiscale2019}
A.~Naso, ``Multiscale analysis of the structure of homogeneous rotating
  turbulence,'' {\em Physical Review Fluids}, vol.~4, no.~2, p.~024609, 2019.

\bibitem{polancoMultiparticle2023}
J.~I. Polanco, S.~Arun, and A.~Naso, ``Multiparticle {{Lagrangian}} statistics
  in homogeneous rotating turbulence,'' {\em Physical Review Fluids}, vol.~8,
  no.~3, p.~034602, 2023.

\bibitem{vanaartrijkSingleparticle2008}
M.~{van Aartrijk}, H.~J.~H. Clercx, and K.~B. Winters, ``Single-particle,
  particle-pair, and multiparticle dispersion of fluid particles in forced
  stably stratified turbulence,'' {\em Physics of Fluids}, vol.~20, no.~2,
  p.~025104, 2008.

\bibitem{xuFlight2014}
H.~Xu, A.~Pumir, G.~Falkovich, E.~Bodenschatz, M.~Shats, H.~Xia, N.~Francois,
  and G.~Boffetta, ``Flight--crash events in turbulence,'' {\em Proceedings of
  the National Academy of Sciences}, vol.~111, no.~21, pp.~7558--7563, 2014.

\bibitem{juchaTimereversalsymmetry2014}
J.~Jucha, H.~Xu, A.~Pumir, and E.~Bodenschatz, ``Time-reversal-symmetry
  {{Breaking}} in {{Turbulence}},'' {\em Physical Review Letters}, vol.~113,
  no.~5, p.~054501, 2014.

\bibitem{braggIrreversibility2018}
A.~D. Bragg, F.~De~Lillo, and G.~Boffetta, ``Irreversibility inversions in
  two-dimensional turbulence,'' {\em Physical Review Fluids}, vol.~3, no.~2,
  p.~024302, 2018.

\bibitem{cheminetEulerian2022}
A.~Cheminet, D.~Geneste, A.~Barlet, Y.~Ostovan, T.~Chaabo, V.~Valori, P.~Debue,
  C.~Cuvier, F.~Daviaud, J.-M. Foucaut, J.-P. Laval, V.~Padilla,
  C.~{Wiertel-Gasquet}, and B.~Dubrulle, ``Eulerian vs {{Lagrangian
  Irreversibility}} in an {{Experimental Turbulent Swirling Flow}},'' {\em
  Physical Review Letters}, vol.~129, no.~12, p.~124501, 2022.

\bibitem{gallonLagrangian2024a}
S.~Gallon, A.~Sozza, F.~Feraco, R.~Marino, and A.~Pumir, ``Lagrangian
  {{Irreversibility}} and {{Energy Exchanges}} in {{Rotating-Stratified
  Turbulent Flows}},'' {\em Physical Review Letters}, vol.~133, no.~2,
  p.~024101, 2024.

\bibitem{scatamacchiaExtreme2012}
R.~Scatamacchia, L.~Biferale, and F.~Toschi, ``Extreme {{Events}} in the
  {{Dispersions}} of {{Two Neighboring Particles Under}} the {{Influence}} of
  {{Fluid Turbulence}},'' {\em Physical Review Letters}, vol.~109, no.~14,
  p.~144501, 2012.

\bibitem{bitaneGeometry2013}
R.~Bitane, H.~Homann, and J.~Bec, ``Geometry and violent events in turbulent
  pair dispersion,'' {\em Journal of Turbulence}, vol.~14, no.~2, pp.~23--45,
  2013.

\bibitem{buariaCharacteristics2015}
D.~Buaria, B.~L. Sawford, and P.~K. Yeung, ``Characteristics of backward and
  forward two-particle relative dispersion in turbulence at different
  {{Reynolds}} numbers,'' {\em Physics of Fluids}, vol.~27, no.~10, p.~105101,
  2015.

\bibitem{buariaExtreme2019}
D.~Buaria, A.~Pumir, E.~Bodenschatz, and P.~K. Yeung, ``Extreme velocity
  gradients in turbulent flows,'' {\em New Journal of Physics}, vol.~21, no.~4,
  p.~043004, 2019.

\bibitem{buariaVorticityStrain2022}
D.~Buaria and A.~Pumir, ``Vorticity-{{Strain Rate Dynamics}} and the {{Smallest
  Scales}} of {{Turbulence}},'' {\em Physical Review Letters}, vol.~128, no.~9,
  p.~094501, 2022.

\bibitem{mininniHybrid2011}
P.~D. Mininni, D.~Rosenberg, R.~Reddy, and A.~Pouquet, ``A hybrid
  {{MPI}}--{{OpenMP}} scheme for scalable parallel pseudospectral computations
  for fluid turbulence,'' {\em Parallel Computing}, vol.~37, no.~6,
  pp.~316--326, 2011.

\bibitem{Rosenberg_2020}
D.~Rosenberg, P.~D. Mininni, R.~Reddy, and A.~Pouquet, ``Gpu parallelization of
  a hybrid pseudospectral geophysical turbulence framework using cuda,'' {\em
  Atmosphere}, vol.~11, no.~2, 2020.

\bibitem{davidsonTurbulence2013}
P.~A. Davidson, {\em Turbulence in {{Rotating}}, {{Stratified}} and
  {{Electrically Conducting Fluids}}}.
\newblock Cambridge: Cambridge University Press, 2013.

\bibitem{Marino_2014}
R.~Marino, P.~D. Mininni, D.~L. Rosenberg, and A.~Pouquet, ``Large-scale
  anisotropy in stably stratified rotating flows,'' {\em Phys. Rev. E},
  vol.~90, p.~023018, Aug 2014.

\bibitem{rileyStratified2008}
J.~J. Riley and E.~Lindborg, ``Stratified {{Turbulence}}: {{A Possible
  Interpretation}} of {{Some Geophysical Turbulence Measurements}},'' {\em
  Journal of the Atmospheric Sciences}, vol.~65, no.~7, pp.~2416--2424, 2008.

\bibitem{zemanNote1994}
O.~Zeman, ``A note on the spectra and decay of rotating homogeneous
  turbulence,'' {\em Physics of Fluids}, vol.~6, no.~10, pp.~3221--3223, 1994.

\bibitem{Pouquet_2019}
A.~Pouquet, D.~Rosenberg, and R.~Marino, ``{Linking dissipation, anisotropy,
  and intermittency in rotating stratified turbulence at the threshold of
  linear shear instabilities},'' {\em Physics of Fluids}, vol.~31, p.~105116,
  10 2019.

\bibitem{iveyDensity2008}
G.~Ivey, K.~Winters, and J.~Koseff, ``Density {{Stratification}},
  {{Turbulence}}, but {{How Much Mixing}}?,'' {\em Annual Review of Fluid
  Mechanics}, vol.~40, no.~1, pp.~169--184, 2008.

\bibitem{batchelorApplication1950}
G.~K. Batchelor, ``The application of the similarity theory of turbulence to
  atmospheric diffusion,'' {\em Quarterly Journal of the Royal Meteorological
  Society}, vol.~76, no.~328, pp.~133--146, 1950.

\bibitem{bitaneTime2012}
R.~Bitane, H.~Homann, and J.~Bec, ``Time scales of turbulent relative
  dispersion,'' {\em Physical Review E}, vol.~86, no.~4, p.~045302, 2012.

\bibitem{pumirTetrahedron2013}
A.~Pumir, E.~Bodenschatz, and H.~Xu, ``Tetrahedron deformation and alignment of
  perceived vorticity and strain in a turbulent flow,'' {\em Physics of
  Fluids}, vol.~25, no.~3, p.~035101, 2013.

\bibitem{juchaTimeSymmetry2015}
J.~Jucha, {\em Time-{{Symmetry Breaking}} in {{Turbulent Multi-Particle
  Dispersion}}}.
\newblock Springer {{Theses}}, Cham: Springer International Publishing, 2015.

\bibitem{finkelQuad1974}
R.~A. Finkel and J.~L. Bentley, ``Quad trees a data structure for retrieval on
  composite keys,'' {\em Acta Informatica}, vol.~4, no.~1, pp.~1--9, 1974.

\bibitem{lillyStratified1983}
D.~K. Lilly, ``Stratified {{Turbulence}} and the {{Mesoscale Variability}} of
  the {{Atmosphere}},'' {\em Journal of the Atmospheric Sciences}, vol.~40,
  no.~3, pp.~749--761, 1983.

\bibitem{billantExperimental2000}
P.~Billant and J.-M. Chomaz, ``Experimental evidence for a new instability of a
  vertical columnar vortex pair in a strongly stratified fluid,'' {\em Journal
  of Fluid Mechanics}, vol.~418, pp.~167--188, 2000.

\bibitem{billantThreedimensional2000}
P.~Billant and J.-M. Chomaz, ``Three-dimensional stability of a vertical
  columnar vortex pair in a stratified fluid,'' {\em Journal of Fluid
  Mechanics}, vol.~419, pp.~65--91, 2000.

\bibitem{biferaleLagrangian2005}
L.~Biferale, G.~Boffetta, A.~Celani, B.~J. Devenish, A.~Lanotte, and F.~Toschi,
  ``Lagrangian statistics of particle pairs in homogeneous isotropic
  turbulence,'' {\em Physics of Fluids}, vol.~17, no.~11, p.~115101, 2005.

\bibitem{betchovInequality1956}
R.~Betchov, ``An inequality concerning the production of vorticity in isotropic
  turbulence,'' {\em Journal of Fluid Mechanics}, vol.~1, no.~5, pp.~497--504,
  1956.

\bibitem{ashurstAlignment1987}
{\relax Wm}.~T. Ashurst, A.~R. Kerstein, R.~M. Kerr, and C.~H. Gibson,
  ``Alignment of vorticity and scalar gradient with strain rate in simulated
  {{Navier}}--{{Stokes}} turbulence,'' {\em The Physics of Fluids}, vol.~30,
  no.~8, pp.~2343--2353, 1987.

\bibitem{moumEnergycontaining1996}
J.~N. Moum, ``Energy-containing scales of turbulence in the ocean
  thermocline,'' {\em Journal of Geophysical Research: Oceans}, vol.~101,
  no.~C6, pp.~14095--14109, 1996.

\bibitem{feraco_18}
F.~Feraco, R.~Marino, A.~Pumir, L.~Primavera, P.~Mininni, A.~Pouquet, and
  D.~Rosenberg, ``Vertical drafts and mixing in stratified turbulence: sharp
  transition with {F}roude number,'' {\em Eur. Phys. Lett.}, vol.~123,
  p.~44002, 2018.
  
\bibitem{feraco_21}
F.~Feraco, R.~Marino, L.~Primavera, A.~Pumir, P.~Mininni, D.~Rosenberg,
  A.~Pouquet, R.~Foldes, E.~L\'ev\^eque, E.~Camporeale, S.~Cerri, H.~C. Asokan,
  J.-L. Chau, J.~Bertoglio, P.~Salizzoni, and M.~Marro, ``Connecting
  large-scale velocity and temperature bursts with small-scale intermittency in
  stratified turbulence,'' {\em Eur. Phys. J.}, vol.~135, 2021.

\bibitem{marino_22}
R.~Marino, F.~Feraco, L.~Primavera, A.~Pumir, A.~Pouquet, D.~Rosenberg, and
  P.~D. Mininni, ``Turbulence generation by large-scale extreme vertical drafts
  and the modulation of local energy dissipation in stably stratified
  geophysical flows,'' {\em Phys. Rev. Fluids}, vol.~7, p.~033801, 2022.

\bibitem{popeTurbulent2000}
S.~B. Pope, {\em Turbulent Flows}.
\newblock Cambridge: Cambridge University Press, 2000.

\bibitem{xuPirouette2011}
H.~Xu, A.~Pumir, and E.~Bodenschatz, ``The pirouette effect in turbulent
  flows,'' {\em Nature Physics}, vol.~7, no.~9, pp.~709--712, 2011.

\end{thebibliography}

\appendix
\section{Identification of Lagrangian Pairs and Tetrahedra}
\begin{figure}[ht]
    \centering    
    \includegraphics[width=\textwidth]{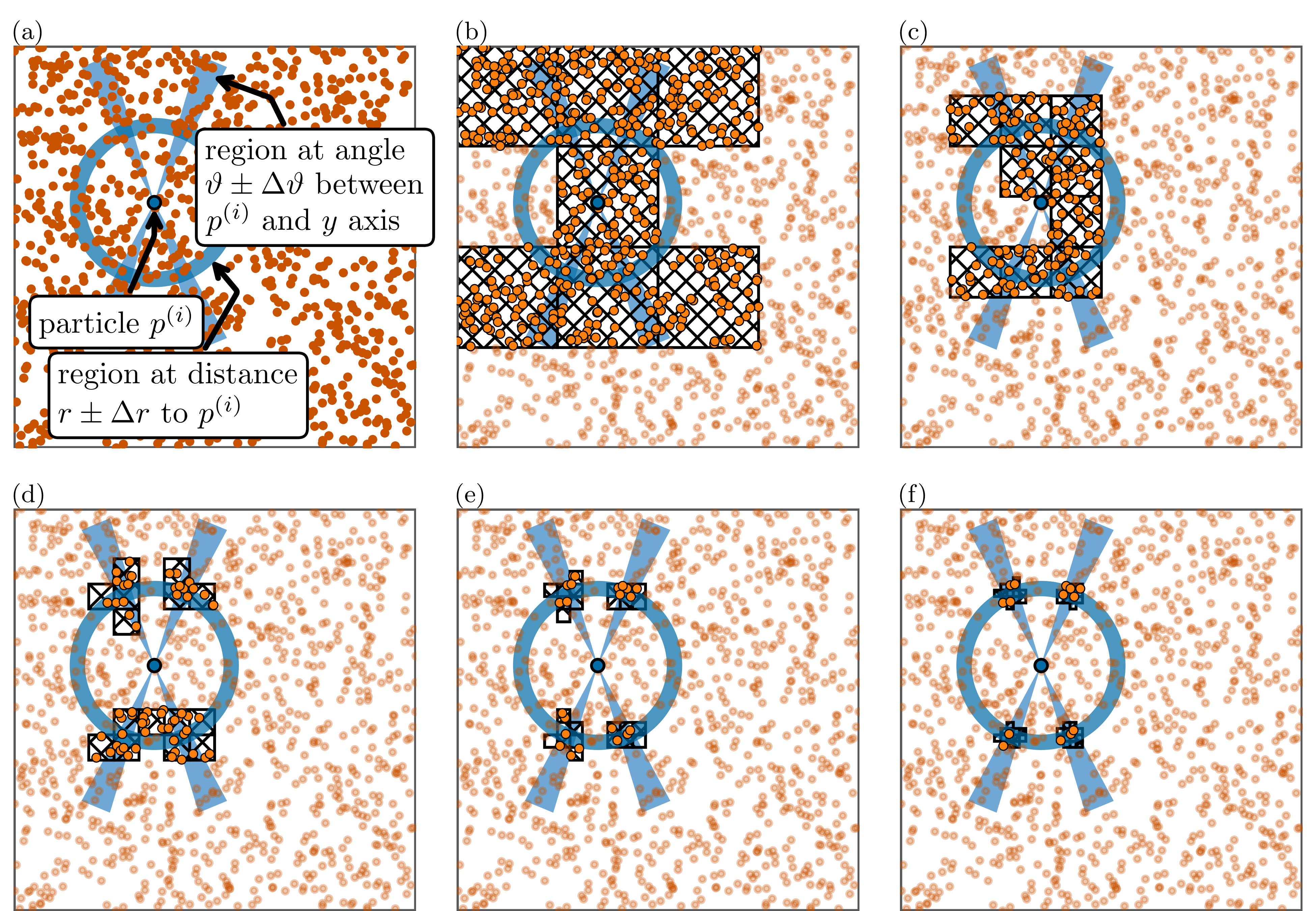}
    \caption{Visualization of the algorithm used to compute pairs of particles that are within a range of separations $r_0 \Delta r_0$ and a predefined orientation parameterized by the angle to the $z$ axis $\vartheta$, up to a tolerance $\Delta \vartheta$ using a quadtree data structure in two dimensions. (a-f) the blue particle with the black edge is the particle $p_1$. (b-f) depiction of the algorithm with increasing level of refinement. The hatched regions are the nodes that intersect with both the spherical shell and the cones. 
    }\label{fig:algorithm}
\end{figure}
The analysis presented below rests in a crucial way on selecting particle pairs initially separated by a specific distance $r_0$, up to a tolerance $\Delta r_0$, and possibly with a separation $\vec{r}_0$ with a specific orientation, i.e. a predefined angle to the $z$-axis 
\begin{equation}
    \cos(\vartheta) = \abs{\frac{\vec{r}_0 \cdot \vec{e}_z}{r_0}} \qq*{,}
\label{eq:angl}
\end{equation}
up to a tolerance $\Delta \vartheta$.Given that an exchange between the two particles merely changes the sign of the scalar product $\vec{r}_0 \cdot \vec{e}_z$, the absolute value in \cref{eq:angl} merely ensures that  $\vartheta$ is in the range $[0,\pi/2]$. The direct approach to this problem would be to compute first all pairwise distances and angles $\vartheta$ to identify the pairs that fulfill the criteria. However, this brute-force approach is prohibitively expensive, the more so as the number of particles increases. To overcome this difficulty, we employ a modified version of the recently proposed algorithm~\cite{gallonLagrangian2024a} using the octree-data structure~\cite{finkelQuad1974} to subdivide the domain into a hierarchy of smaller and smaller cubes. The original domain is subdivided into eight equal-sized sub-cubes. The original cube is referred to as root node and the sub-cubes as its children. Each child node is recursively subdivided for a fixed number of refinements $m$, leading to nodes at the smallest level denoted as leaf nodes. The particles are then assigned to the leaf nodes, based on their position. To chose the number of refinements $m$, we found that the proposed method works best with at least 10 particles per leaf node.\par
The algorithm to select particle pairs then proceeds as follows. We start by picking a particle $p_1$. All particles $p_2$ that have both at a separation in $[r_0 - \Delta r_0, r_0 + \Delta r_0]$ and an angle to the $z$-axis in $[\vartheta - \Delta \vartheta, \vartheta + \Delta \vartheta]$ are all located in the intersection of a spherical shell and two cones, with rotational symmetry around the $z$-axis, as sketched in \cref{fig:algorithm} panel (a). Similarly to the previous work~\cite{gallonLagrangian2024a}, the pair partner $p_2$ are found by traversing the tree. We start at the leaf node containing $p_1$ and going up the tree, until we arrive at a node, whose side length $l_i$ fulfills $l_i/2 < r_0 + \Delta r_0 < l_i$. All particles that are within the spherical shell $[r_0 - \Delta r_0, r_0 + \Delta r_0]$ must be either in this node or in one of its neighbors. We then verify that each of the candidate nodes intersects with the two rotated cones. If a node does not intersect with the cones, we can discard it altogether with its children nodes, see \cref{fig:algorithm} panel (b). We then descent the tree until we reach the leaf nodes. In each step, we check if the nodes intersects with the cones and the spherical shell, that is if the particles in the node are neither too far away or too close and have may have the right orientation. If a node does not fulfill the criteria, we can discard it and all its children nodes. This decent is depicted in \cref{fig:algorithm} panels (c-f). Finally, when arriving at the leaf nodes, we directly compute the pairwise distances and angles and identify the pairs that fulfill the criteria. \par
The extension of this algorithm to finding groups of four equidistant particles is straight-forward. To this end, we start by selecting a particle $p_1$ and identify a particle $p_2$ that is within the desired distance range using the algorithm sketched above. We then search the tree similarly as before to identify a third particle $p_3$ that is within the correct distance-range to both $p_1$ and $p_2$. Once such a particle is found, we repeat the procedure to find a particle $p_4$ that is within the correct distance range to $p_1$, $p_2$ and $p_3$.
\section{Prediction of the horizontal and vertical second-order structure functions for HIT}\label{apx:B}
To predict the angle-dependence of the vertical and horizontal second-order structure functions defined in \cref{eq:struc_fns1,eq:struc_fns2}, we consider the covariance tensor of the velocity differences
\begin{equation}
    D_{ij}(\vec{x}^{(1)},\vec{x}^{(2)},t) = \average{\qty[u_i(\vec{x}^{(2)},t)-u_i(\vec{x}^{(1)},t)]\qty[u_j(\vec{x}^{(2)},t)-u_j(\vec{x}^{(1)},t)] } \label{eq:covTensorD}
\end{equation}
For homogeneous and isotropic flows, $D_{ij}$ can be simplified to~\cite{moninStatistical1975,popeTurbulent2000}
\begin{equation}
    D_{ij}(\vec{r},t) = D_{NN}(r,t) \delta_{ij} + \qty(D_{LL}(r,t)-D_{NN}(r,t))\frac{r_i r_j}{r^2} \qc \label{eq:covTensorIsotropic}
\end{equation}
with the two scalar functions $D_{LL}(r,t)$ and  $D_{NN}(r,t)$ being the longitudinal and transversal structure-function contributions. Moreover, the transversal and longitudinal components for homogeneous and isotropic flows are related by 
\begin{equation}
    D_{NN}(r,t) = D_{LL}(r,t) + \frac{1}{2} r \pdv{r} D_{LL}(r,t) \qq*{.} \label{eq:apx1}
\end{equation}
We now insert the prediction of the K41 theory of turbulence~\cite{moninStatistical1975,popeTurbulent2000} $D_{LL}(r,t)\propto r^{2/3}$ in the inertial range, i.e., $\eta \ll r \ll L$, and $D_{LL}(r,t)\propto r^{2}$ in the dissipative range, i.e., $r \ll \eta$ in \cref{eq:apx1} and insert the resulting expression for $D_{NN}$ in \cref{eq:covTensorIsotropic}. With $\cos^2(\vartheta)= r_3r_3/r^2$ and $\sin^2(\vartheta)=(r_1r_1 + r_2 r_2)/r^2$, we rewrite the second order vertical structure function as 
\begin{align}
    S_{2,V}(r,\vartheta)&=D_{33}(\vec{r},t) =
    \begin{cases}
        D_{LL}(r,t)\qty[2 - \cos^2(\vartheta)] & \qq{for} r \ll \eta \qq*{,} \\
        \dfrac{D_{LL}(r,t)}{3}\qty[4 - \cos^2(\vartheta)] & \qq{for} \eta \ll r \ll L \qq*{,} 
    \end{cases}  \label{eq:stre_funct_vert}   \\
    S_{2,H}(r,\vartheta)&=D_{11}(\vec{r},t)+D_{22}(\vec{r},t) =
    \begin{cases}
        D_{LL}(r,t)\qty[4 - \sin^2(\vartheta)] & \qq{for} r \ll \eta \qq*{.} \\
        \dfrac{D_{LL}(r,t)}{3}\qty[8 - \sin^2(\vartheta)] & \qq{for} \eta \ll r \ll L \qq*{.}  \label{eq:stre_funct_hor} 
    \end{cases} 
\end{align}
The averages over all orientations are then given by
\begin{equation}
\bar{S}_{2,V}(r) = \int_{0}^{\frac{\pi}{2}} \dd{\vartheta} S_{2,V}(r,\vartheta)\sin(\vartheta) =
\begin{cases}
    \frac{5}{3}  D_{LL}(r,t)& \qq{for} r \ll \eta  \qq*{,} \\
    \frac{11}{9} D_{LL}(r,t)& \qq{for} \eta \ll r \ll L \qq*{.} 
\end{cases} \label{eq:av_stre_funct_vert}
\end{equation}
and
\begin{equation}
    \bar{S}_{2,H}(r) = \int_{0}^{\frac{\pi}{2}} \dd{\vartheta} S_{2,H}(r,\vartheta)\sin(\vartheta) =  
    \begin{cases}
    \frac{10}{3}D_{LL}(r,t) & \qq{for} r \ll \eta \qq*{,}\\
    \frac{22}{9}D_{LL}(r,t)& \qq{for} \eta \ll r \ll L \qq*{.}
    \end{cases}
\label{eq:av_stre_funct_hor}
\end{equation}
\section{Alignment of the perceived rate-of-strain tensor's eigenvectors}\label{AppendixB}
\begin{figure}[t]
    \centering
    \includegraphics[width=\textwidth]{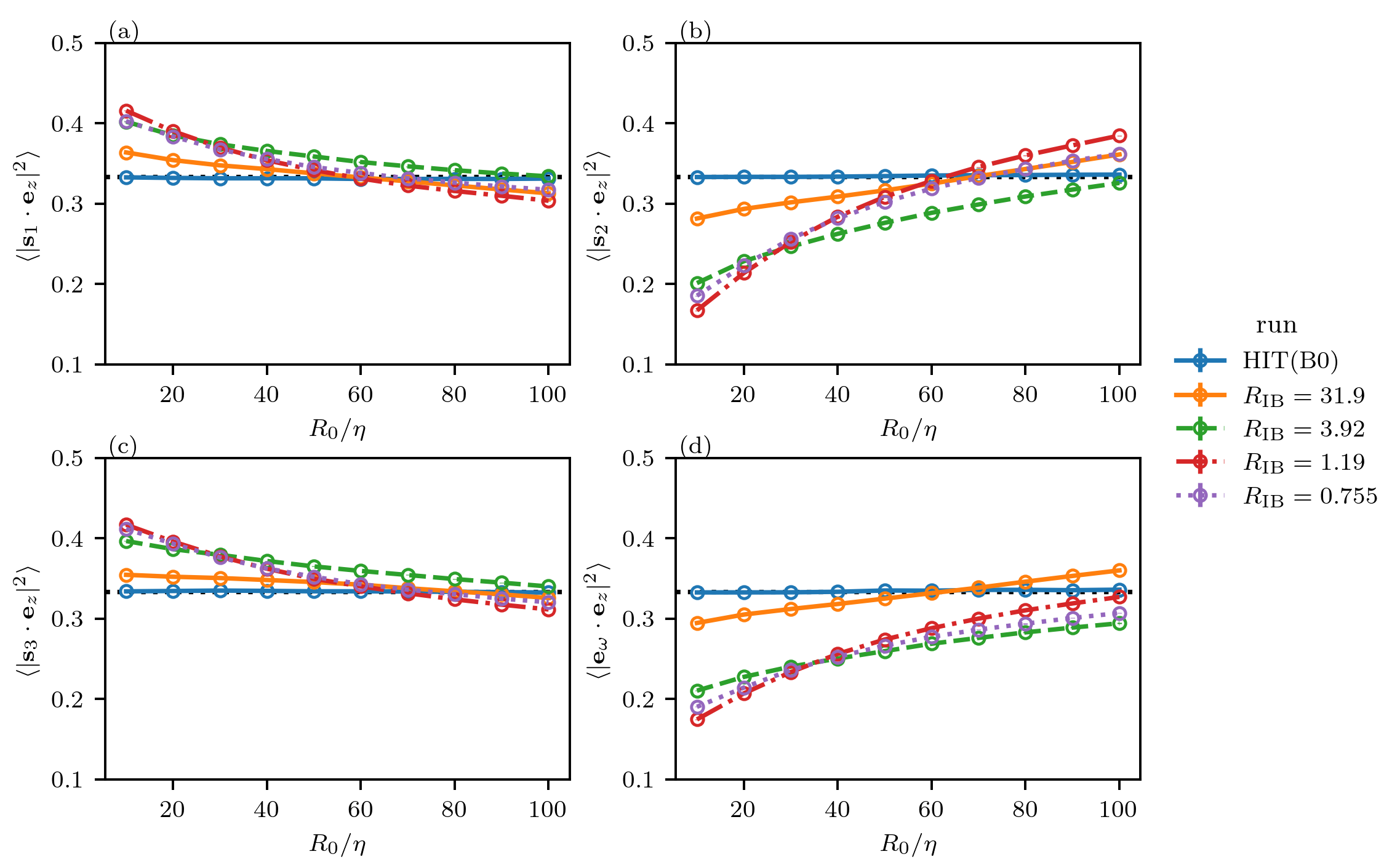}
    \caption{Alignment between the eigenvalues of strain, $\mathbf{s}_i$, and $\vec{e}_z$, $\langle (\vec{s}_i \cdot \vec{e}_z)^2 \rangle$, as a function of the size $R_0$ of the tetrahedra, for $i=1$ (a), $i=2$ (b) and $i=3$ (c). }
    \label{fig:aignt_strain}
\end{figure}

In a flow, in the presence of stratification and rotation, the vertical direction $\vec{e}_z$ plays a particular role. As a result, one observes some alignment between the eigenvectors of the strain tensor, $\vec{s}_i$, corresponding to the eigenvalues $s_i$, in decreasing order. This is illustrated by \cref{fig:aignt_strain}, which reveals a mild alignment of $\vec{s}_1$ and $\vec{s}_3$ with $\vec{e}_z$, and a mild anti-alignment between $\vec{s}_2$ and $\vec{e}_z$.  This observation can be explained using a heuristic argument:
For HIT, the eigenvector corresponding to the  intermediate eigenvalue of the perceived rate-of-strain tensor  is preferentially aligned to the vorticity~\cite{xuPirouette2011,pumirTetrahedron2013}, which we confirm for rotating stratified turbulence when comparing the panels (b) and (d) in \cref{fig:aignt_strain}. In the setup we consider here, where $N/f=5$, stratification causes a generation of horizontal vorticity, as we can see when taking the curl of the Boussinesq equation \cref{eq:Bous_u}, neglecting viscous contributions and linearizing the equations. We then obtain
\begin{equation}
  \pdv{t} \boldsymbol{\omega} \approx f \pdv{z} \vec{u} + N \vec{e}_z \cross \grad \theta \qq*{.}
\end{equation}
Stratification therefore generates vorticity perpendicular to $ \vec{e}_z$, i.e., in the horizontal plane. The intermediate rate-of-strain eigenvector then aligns with the vorticity and is therefore slightly preferential perpendicular to $ \vec{e}_z$. For geometric reasons, the dealignment of the intermediate eigenvector, results in a weak alignment of both the first and third eigenvector that is visible in \cref{fig:aignt_strain}. We remark that this argumentation is heuristic and remains to be understood in a deeper way. Moreover, we note that this observation is contrary to what happens in purely rotating turbulent flows~\cite{nasoMultiscale2019}.
 
\end{document}